\documentclass[a4paper, aps, prd, showpacs, superscriptaddress, nofootinbib, twocolumn]{revtex4-1}
\usepackage{graphicx, color}
\usepackage{amsmath, amssymb, amscd, latexsym, bm, braket}
\usepackage[colorlinks=true,pdfstartview=FitV,linkcolor=blue,citecolor=magenta,urlcolor=blue,bookmarks=true,bookmarksnumbered=true]{hyperref}

\begin{document}
\title{Real-time dynamics of axion particle production due to spontaneous decay of a coherent axion field}

\author{Xu-Guang Huang}
\email{{\tt huangxuguang@fudan.edu.cn}}
\address{Department of Physics and Center for Field Theory and Particle Physics, Fudan University, Shanghai, 200433, China }
\address{Key Laboratory of Nuclear Physics and Ion-beam Application (MOE), Fudan University, Shanghai 200433, China }

\author{Dmitri E.~Kharzeev}
\email{{\tt dmitri.kharzeev@stonybrook.edu}}
\address{Department of Physics and Astronomy, Stony Brook University, Stony Brook, New York 11794-3800, USA}
\address{Department of Physics, Brookhaven National Laboratory, Upton, New York 11973-5000, USA}
\address{RIKEN-BNL Research Center, Brookhaven National Laboratory, Upton, New York 11973-5000, USA}

\author{Hidetoshi Taya}
\email{{\tt h_taya@fudan.edu.cn}}
\address{Department of Physics and Center for Field Theory and Particle Physics, Fudan University, Shanghai, 200433, China }

\date{\today}

\begin{abstract}
We show that the coherent axion field spontaneously decays by emitting axion particles via quantum tunneling if axion potential has more than one minimum.  By employing the $\hbar$-expansion and mean field approximation, we develop a formalism to trace the real-time dynamics of the axion particle production including its backreaction to the coherent axion field. We also present numerical results for the time evolution of various physical quantities including the strength of the coherent axion field, phase-space density of produced axion particles, energy density, and pressure. Phenomenological implications of our results are also discussed.
\end{abstract}

\maketitle

\section{Introduction}

The Peccei-Quinn mechanism is one of the most compelling resolutions to the strong CP problem \cite{pec77a, pec77b}.  The idea is that the $\theta$-angle of quantum chromodynamics (QCD) is promoted to be a dynamical field $\theta \to \hat{\phi}$ by introducing an axial U(1) symmetry (Peccei-Quinn symmetry).  Low energy dynamics of the field $\hat{\phi}$ is determined by the topological structure of the QCD vacuum, because of which an effective potential for the field has a stable minimum at $\braket{\hat{\phi}} = 0$ \cite{vaf84}.  Therefore, the vacuum expectation value of the field $\braket{\hat{\phi}}$ dynamically evolves toward the minimum $\braket{\hat{\phi}} = 0$ no matter what the initial expectation value is, and the strong CP problem is resolved.

The Peccei-Quinn mechanism predicts the existence of a new particle, called ``axion,'' as quantum of the newly introduced field $\hat{\phi}$ \cite{wei78, wil78} (for reviews, see Refs.~\cite{kim87, sik08, pec08, kim10, mar16}).  Axion particles\footnote{We distinguish ``coherent axion field'' and ``axion particles.''  The former is a classical field which is defined as a vacuum expectation value of the axion field operator $\braket{\hat{\phi}}$, while the latter is incoherent quantum fluctuations on top of the classical field, $\hat{\varphi} \equiv \hat{\phi} - \braket{\hat{\phi}}$.  Note that we do not distinguish ``coherent axion field'' and ``axion condensate'' because in this paper both are defined as the vacuum expectation value $\braket{\hat{\phi}}$.  } may be produced, for example, in stars (e.g. the sun) via perturbative processes such as Primakoff effect ($\gamma + \gamma \to {\rm axion}$) \cite{dic78, fuk82a, fuk82b, raf86}, Compton scattering ($\gamma + e \to e + {\rm axion}$) \cite{fuk82a, fuk82b, raf86}, bremsstrahlung ($e +  Z \to e + Z +{\rm axion}$) \cite{kra84, raf86}, and axio-recombination ($e +  Z \to  Z^- +{\rm axion}$) \cite{dim86}.  Many other possible production mechanisms are also proposed in the literatures, which include, to name a few, thermal production in the early Universe \cite{tur87, mas02, gra11, sal14}, cosmic-ray induced emission from axion condensate \cite{hua19}, and parametric resonance \cite{pat99, ray18, sod18, kia18, hay19}.  Unfortunately, in spite of the great experimental/observational efforts over the past four decades, axion particles have not been detected yet (see \cite{ros00, bra03, asz06, gra15} for experimental reviews).  Nevertheless, the past experiments/observations constrained axion parameters such as the mass $m$ and the decay constant $f_a$, and the coupling constants to standard model particles (e.g. photon, electron, nucleon) which are very small.

In the present paper, we propose a novel axion particle production mechanism by the spontaneous decay of coherent axion field induced by quantum tunneling.  Namely, we consider a situation in which coherent axion field is excited $\braket{\hat{\phi}} \neq 0$ at some instant.  Such a coherent excitation may be seeded by, for example, the misalignment mechanism \cite{pre83, abb83, mic83}, QCD sphalerons (e.g. the cosmological QCD phase transition, heavy-ion collisions), and classical electromagnetic fields with nonvanishing ${\bm E}\cdot{\bm B}$ (e.g. solar flares).  As the QCD vacuum has $2\pi$-periodicity in terms of the $\theta$-angle, the axion potential would have infinitely degenerate minima in terms of the field $\hat{\phi}$.  This implies that if one considers quantum fluctuations on top of the coherent field, $\hat{\varphi} \equiv \hat{\phi} - \braket{\hat{\phi}}$, the coherent field initially peaked at some value $\braket{\hat{\phi}} \neq 0$ would be delocalized because of quantum tunneling between different minima and the quantum fluctuation $\braket{|\hat{\varphi}|^2}$ grows.  This is equivalent to saying that the coherent axion field is unstable against quantum fluctuations because of quantum tunneling, and it decays spontaneously by emitting axion particles.  Notice that the existence of many minima is important here.  For a single-minimum potential, quantum fluctuations cannot grow and the coherent axion field is stable because that quantum tunneling does not take place.  The physical mechanism is essentially the same as the particle production mechanism in the presence of classical field such as Hawking radiation \cite{haw74, haw75}, Schwinger mechanism \cite{sau31, hei36, sch51}, and Landau-Zener transition \cite{lan32, zen32, stu32, maj32}, in which classical field becomes unstable and decays spontaneously by emitting particles because of quantum tunneling.

We shall discuss the aforementioned axion production mechanism based on quantum field theory (QFT).  Field theoretical study of real-time dynamics of particle production in the presence of classical field has been developed significantly over the past decades.  In particular, in the context of the Schwinger mechanism, not only the dynamical evolution of the production number, but also the backreaction by the produced particles was elegantly formulated based on QFT within mean field approximation by using a proper ultraviolet (UV) regularization scheme \cite{coo89, klu91, klu92, klu93, tan09}.  Inclusion of backreaction is essential for the energy conservation of the system and for describing the decay dynamics of coherent axion field.  The mean field approximation neglects scattering between produced particles (which is responsible for mode-couplings).  Therefore, the use of the mean field approximation is justified when coupling constants are sufficiently small \cite{mue02}, which is actually the case for axion.  More precisely, the product between the phase-space density $f$ and the coupling constant $g$ determines whether the scattering effects beyond the mean field approximation are important.  This is intuitively because scattering takes place more frequently if there are more particles.  For the axion self-coupling constant $g \sim \chi/f_a^4 \lesssim 10^{-40}$, where $\chi \sim \Lambda_{\rm QCD}^4$ is the topological susceptibility of QCD, huge phase-space density $f \gg 1/g \sim 10^{40}$ is required for the scattering effects to be important.  For such huge phase-space density, the axion dynamics would be analyzed classically within, for example, the lattice simulation technique \cite{ami14}.  In the present study, however, we shall explicitly show that the phase-space density of produced axions via the present axion production mechanism is far from dense $f \gg\!\!\!\!\!\!/ \ 1/g$, and therefore our mean field treatment is justified.

The present paper is organized as follows: In Sec.~\ref{sec2}, we formulate the axion production mechanism based on QFT within the $\hbar$-expansion and mean field approximation.  In Sec.~\ref{sec3}, we present numerical results based on our formulation and discuss the real-time dynamics of the axion particle production mechanism quantitatively.  Section~\ref{sec4} is devoted to summary and discussion.

Throughout this paper, we use the mostly minus metric $g_{\mu\nu} = {\rm diag}(+1,-1,-1,-1)$, and work in the Heisenberg picture.  We adopt the natural units, and $\hbar$ is not written explicitly in the following as long as it is not confusing.

\section{Formalism} \label{sec2}

In this section, we explain the theoretical basis of this study.  In Sec.~\ref{sec2a}, in order to describe the axion particle production mechanism from coherent axion field, we employ the $\hbar$-expansion and mean field approximation to derive the equations of motion for the coherent axion field and dynamical axion field on top of the coherent field.  In Sec.~\ref{sec2b}, we explain how to formulate axion particle production with the derived equations of motion.  In this study, we adopt the adiabatic particle picture, and canonically quantize the dynamical axion field at each instant of time.  We, then, directly evaluate the vacuum expectation value of the axion number operator, and see that it becomes nonzero if axion potential has more than one minimum.  In Sec.~\ref{sec2c}, we discuss the energy balance of the system to check the consistency of our formulation and to see the importance of the backreaction for the energy conservation and describing the decay dynamics of the coherent axion field.

\subsection{Equations of motion} \label{sec2a}

We consider an axion Lagrangian given by
\begin{align}
	{\mathcal L} = \frac{1}{2} (\partial_{\mu} \hat{\phi}) (\partial^{\mu} \hat{\phi}) - V[\hat{\phi}], \label{eq1}
\end{align}
where $\hat{\phi}$ is the axion field and $V$ is the axion potential.  In this paper, we assume for simplicity that the system is not expanding and is homogeneous in space.

Since we are interested in axion particle production from coherent axion field, it is convenient to separate the total axion field $\hat{\phi}$ into dynamical axion field $\hat{\varphi}$ and coherent axion field $\bar{\phi}$ as
\begin{align}
	\hat{\phi} = \bar{\phi} + \hat{\varphi}.
\end{align}
Here, the coherent field $\bar{\phi}$ is defined as a classical expectation value of the total field operator $\hat{\phi}$ for a given in-state $\ket{\rm in}$ as
\begin{align}
	\bar{\phi} \equiv \braket{{\rm in}| \hat{\phi} | {\rm in}} .
\end{align}

The coherent axion field is a classical quantity and we assume that it is of order $\bar{\phi} = {\mathcal O}(1)$.  On the other hand, the dynamical axion field is a quantum fluctuation, so we assume $\hat{\varphi} = {\mathcal O}(\hbar^{1/2})$.  With this power counting rule, one may expand the Lagrangian (\ref{eq1}) in terms of $\hbar$ (or $\hat{\varphi}$) as
\begin{align}
	{\mathcal L}
		&= 	\frac{1}{2} (\partial_{\mu} \bar{\phi}) (\partial^{\mu} \bar{\phi}) - V[\bar{\phi}] \nonumber\\
		&\quad	+ (\partial_{\mu} \bar{\phi}) (\partial^{\mu} \hat{\varphi}) - V'[\bar{\phi}] \hat{\varphi} \nonumber\\
		&\quad	+ \frac{1}{2} (\partial_{\mu} \hat{\varphi}) (\partial^{\mu} \hat{\varphi}) - \frac{1}{2}V''[\bar{\phi}] \hat{\varphi}^2 \nonumber\\
		&\quad	+ {\mathcal O}(\hbar^{3/2}).  \label{eq4}
\end{align}

From Eq.~(\ref{eq4}), one can derive the equation of motion up to ${\mathcal O}(\hbar^{3/2})$ as
\begin{align}
	0	&=		\partial_{\mu} \partial^{\mu} \bar{\phi} + V'[\bar{\phi}] \nonumber\\
		&\quad	+ \left[ \partial_{\mu}\partial^{\mu} + V''[\bar{\phi}] \right] \hat{\varphi} \nonumber\\
		&\quad	+ \frac{1}{2} V'''[\bar{\phi}]\hat{\varphi}^2 \nonumber\\
		&\quad	+ {\mathcal O}(\hbar^{3/2}).  \label{eq5}
\end{align}
In the following, we ignore the ${\mathcal O}(\hbar^{3/2})$ term\footnote{For a single-minimum potential with higher order nonlinear terms, higher order $\hbar$-corrections become important.  For example, if we consider a quartic potential $V \propto  \phi^4$, the $\hbar^2$-correction effectively gives a mass term $\sim \braket{\hat{\varphi}^2}\hat{\varphi}^2$ for the Lagrangian of the dynamical field.  This term gives a strong negative feedback to the dynamical field to grow, i.e., axion particle production is strongly suppressed.  Physically, this is because a single-minimum potential does not permit quantum tunneling and the dynamical field cannot go far away from the minimum.}.  By taking the in-in expectation value of Eq.~(\ref{eq5}) (mean field approximation), one obtains equations of motion for the coherent field $\bar{\phi}$ and the dynamical field $\hat{\varphi}$ as
\begin{subequations}
\begin{align}
	0 &= \partial_{\mu} \partial^{\mu} \bar{\phi} + V'[\bar{\phi}] + \frac{1}{2} V'''[\bar{\phi}] \braket{{\rm in}| : \hat{\varphi}^2 : |{\rm in}} ,  \label{eq6a} \\
	0 &=  \left[ \partial_{\mu}\partial^{\mu} + V''[\bar{\phi}] \right] \hat{\varphi} .   \label{eq6b}
\end{align}
\end{subequations}
In computing the third term, one has to be careful about how to regularize the UV-divergence in the two-point function $\braket{{\rm in}| \hat{\varphi}^2 |{\rm in}} \to \braket{{\rm in}|: \hat{\varphi}^2 : |{\rm in}}$.  Note that we will give a precise definition of the regularization $\braket{: \hat{\bullet} :}$ later in Sec.~\ref{sec2C1} (see Eq.~(\ref{eq23})) based on the adiabatic regularization scheme \cite{bir82}, which can be understood as an analog of the Pauli-Villars regularization scheme in usual QFT in equilibrium.  For the moment, we just use $\braket{: \hat{\bullet} :}$ to mean a regularized value of $\braket{ \hat{\bullet} }$.  We also note that $\hat{\varphi}^2 - \braket{{\rm in}|: \varphi^2 :| {\rm in}}$ is higher order in the $\hbar$-expansion compared to $\hat{\varphi}^2$ and $\braket{{\rm in}|: \varphi^2 :| {\rm in}}$.  This is because the expectation value of the difference $\braket{{\rm in}|: \hat{\varphi}^2 - \braket{{\rm in}|: \varphi^2 :| {\rm in}}:| {\rm in} } $ is vanishing, and it can give nonvanishing contribution after, at least, it is squared $\braket{{\rm in}|: (\hat{\varphi}^2 - \braket{{\rm in}|: \varphi^2 :| {\rm in}})^2  :| {\rm in} } \neq 0$, which is higher in $\hbar$.  Therefore, as long as the $\hbar$-expansion is justified, one can safely neglect the term $\hat{\varphi}^2 - \braket{{\rm in}|: \varphi^2 :| {\rm in}}$ in Eq.~(\ref{eq6b}).

An important point in Eqs.~(\ref{eq6a}) and (\ref{eq6b}) is that the dynamical axion field and the coherent field are coupled with each other when the axion potential $V$ is higher than quadratic in $\hat{\phi}$, i.e., the axion potential $V$ has more than one minimum.  This point is essential for the axion particle production as we shall discuss in the next section \ref{sec2b}.

The third term in Eq.~(\ref{eq6a}) represents backreaction of axion particle production to the coherent field.  This term is higher in the $\hbar$-expansion than the other terms, however, one should not dismiss this term because it is crucially important for the energy conservation of the system.  Indeed, we shall explicitly see later that without this term axion particles are endlessly created so that the energy eventually diverges (see Sec.~\ref{sec3b}).

Before passing, it is instructive to discuss more about the validity of the $\hbar$-expansion in terms of the size of the phase-space density $f$ and the axion self-coupling constant $g \equiv m^2/f_a^2 \sim \chi/f_a^4$, where $m$ is the axion mass and $f_a$ is the axion decay constant.  As will be shown later (see Eq.~(\ref{eq43})), the two-point function $\braket{{\rm in}|: \hat{\varphi}^2 :|{\rm in}}$ and the total number of produced axions $N$ are related with each other as $\braket{{\rm in}|: \hat{\varphi}^2 :|{\rm in}} \sim N/(m V)$.  Then, by assuming that the typical momentum scale of produced axions is given by the axion mass scale $m$ as $\int d^3{\bm p} \sim (4\pi/3)m^3 \sim m^3$ (which is actually consistent with our numerical simulations; see Figs.~\ref{fig4} and \ref{fig9}), one may estimate the magnitude of $\hat{\varphi}$ in terms of the phase-space density $f \equiv (2\pi)^3 d^6N/d{\bm x}^3 d{\bm p}^3 \sim 10^2 \times N/(m^3 V)$ as
\begin{align}
	{\mathcal O}(\hat{\varphi}) = {\mathcal O}(\sqrt{ \braket{{\rm in}|: \hat{\varphi}^2 :|{\rm in}} }) = {\mathcal O}( f_a \times \sqrt{ 10^{-2} \times gf }).  
\end{align}
By noting that $\hat{\varphi}^n$ is always accompanied by $V^{(n)} \sim g f_a^{-n}$ in the $\hbar$-expansion in Eq.~(\ref{eq4}), one may understand that the $\hbar$-expansion is equivalent to an expansion in terms of $gf$.  Therefore, the $\hbar$-expansion (\ref{eq4}) is justified as long as the condition, 
\begin{align}
	gf \lesssim 100,
\end{align}
is satisfied.  In other words, the $\hbar$-expansion becomes invalid if the phase-space density becomes extremely huge as $f \gg 100/g \gtrsim 10^{42}$.  We shall show explicitly that this is not the case for our axion production mechanism (see Figs.~\ref{fig8} and \ref{fig9}).

\subsection{Axion particle production} \label{sec2b}

We shall solve the coupled equations (\ref{eq6a}) and (\ref{eq6b}) as an initial value problem to trace the dynamical evolution of $\bar{\phi}$ and $\hat{\varphi}$.  In this subsection, we first explain how to obtain information of (or define) axion particles from the field operator $\hat{\varphi}$ on the basis of the adiabatic particle picture \cite{bir82}.  We, then, show that dynamical axion particles are spontaneously produced from the coherent axion field $\bar{\phi} \neq 0$ if the axion potential has more than one minimum.

\subsubsection{Adiabatic particle picture}

In order to explain the adiabatic particle picture, let us first recall how to define ``particle'' in a system in equilibrium with the standard canonical quantization procedure.  First, we expand the dynamical field operator $\hat{\varphi}$ in terms of a mode function $\varphi^{\rm (eq)}_{\bm p}$ as
\begin{align}
	\hat{\varphi}(x) = \int d^3{\bm p} \left[ \varphi^{\rm (eq)}_{\bm p}(t) \hat{a}^{\rm (eq)}_{\bm p} + \varphi^{{\rm (eq)}*}_{-{\bm p}}(t) \hat{a}_{-{\bm p}}^{{\rm (eq)}\dagger}  \right] \frac{{\rm e}^{i{\bm p}\cdot{\bm x}}}{(2\pi)^{3/2}},
\end{align}
where we normalized the mode function $\varphi^{\rm (eq)}_{\bm p}$ as
\begin{align}
	1 &= + i \varphi_{\bm p}^{{\rm (eq)}*} \overset{\leftrightarrow}{\partial}_{t} \varphi_{\bm p}^{\rm (eq)},  \label{eq8}
\end{align}
where $\overset{\leftrightarrow}{\partial} \equiv \overset{\rightarrow}{\partial} - \overset{\leftarrow}{\partial}$.  In general, the choice of $\varphi^{\rm (eq)}_{\bm p}$ is not unique.  That is, one can choose arbitrary $\varphi^{\rm (eq)}_{\bm p}$ as long as it satisfies the equation of motion.  However, an equilibrium system should be invariant with respect to time translation.  Therefore, it is natural to respect this invariance, and so one can uniquely identify the mode function $\varphi^{\rm (eq)}_{\bm p}$ as an eigen-function of the invariance (i.e., the plane wave) as
\begin{align}
	\varphi^{\rm (eq)}_{\bm p} = \frac{ 1}{\sqrt{2\sqrt{m^2 + {\bm p}^2}}} {\rm e}^{- i \sqrt{m^2+{\bm p}^2}t},
\end{align}
where axion mass $m$ is defined as the value of $\sqrt{V''}$ in equilibrium and is a positive constant; otherwise the system cannot be equilibrated.  After the mode expansion, we impose the canonical commutation relation onto $\hat{\varphi}$, which in turn quantizes $\hat{a}^{\rm (eq)}_{\bm p}$ to define particles.  Because of the normalization condition (\ref{eq8}), the annihilation operator satisfies
\begin{align}
	\delta^3({\bm p}-{\bm p}') = [ \hat{a}^{\rm (eq)}_{\bm p}, \hat{a}^{{\rm (eq)}\dagger}_{{\bm p}'} ].  \label{eq10}
\end{align}
Also, the annihilation operator $\hat{a}^{\rm (eq)}_{\bm p}$ can be expressed in terms of the mode function $\varphi^{\rm (eq)}_{\bm p}$ as
\begin{align}
	\hat{a}^{\rm (eq)}_{\bm p} =  + i \int d^3{\bm x} \frac{{\rm e}^{-i{\bm p}\cdot{\bm x}}}{(2\pi)^{3/2}} \varphi_{\bm p}^{{\rm (eq)}*} \overset{\leftrightarrow}{\partial}_{t} \hat{\varphi}.  \label{eq11}
\end{align}
It is evident from this expression that a different mode function defines a different annihilation operator.  We stress that, in equilibrium, one can uniquely define the annihilation operator $\hat{a}^{\rm (eq)}_{\bm p}$ because one can uniquely identify the mode function $\varphi^{\rm (eq)}_{\bm p}$ thanks to the time-translational invariance.  In nonequilibrium, however, the time-translational invariance is explicitly broken so that one has to adopt (or assume) another guiding principle to identify a mode function to define an annihilation operator.

In the adiabatic particle picture, we assume that a given nonequilibrium system is sufficiently adiabatic\footnote{The ``adiabaticity'' is guaranteed if $\hbar$ times time derivatives of $\omega_{\bm p}$ divided by appropriate powers of $\omega_{\bm p}$ (i.e., $\hbar^n \omega_{\bm p}^{(n)}/\omega_{\bm p}^{n+1}$ ($n\in {\mathbb N}$)) is small because a time derivative appears with $\hbar$ in the equation of motion.  This condition is necessary to justify the use of the lowest order WKB solution (\ref{eq14}) as the adiabatic mode function $\varphi^{\rm (ad)}_{\bm p}$ to define the adiabatic annihilation operator $\hat{a}_{\bm p}^{\rm (ad)}$.  Higher order WKB solutions could be significant if $\hbar$ is large or $\omega_{\bm p}^{(n)}$ becomes very large such that it compensates the smallness of $\hbar$.  If this is the case, it could be more appropriate to use a higher order WKB solution as the adiabatic mode function; although this is not the case in the present problem because we are always assuming the smallness of $\hbar$ throughout our formulation and $\omega_{\bm p}^{(n)}$ remains small as long as the typical energy scale of axion condensate is characterized by the QCD energy scale, which is a reasonable assumption in axion phenomenology.  A different adiabatic mode function would not only affect the intermediate particle number, but also would change the time evolution of two-point functions because we adopt the adiabatic regularization scheme to eliminate UV-divergence in two-point functions, whose subtraction does depend on the choice of the adiabatic mode function (see Sec.~\ref{sec2c}).  The dependence of the truncation order of the WKB expansion for the adiabatic mode function to the intermediate particle number was previously discussed, for example, in Refs.~\cite{dab16, hab99}.  In particular, Ref.~\cite{dab16} considered an analytically solvable model of the Schwinger mechanism, and proposed that there exists an optimal truncation order, around which effects of the truncation order become less significant and universal particle number close to the exact one can be obtained.  Therefore, it could be desirable for large $\hbar$ and/or $\omega_{\bm p}^{(n)}$ to consider the optimal truncation order for the adiabatic mode function.  However, it is, in general, impossible to determine the optimal order without knowing the exact solution.  Also, it is known that higher WKB solutions can lead to unphysical time evolution of a system (cf. backreaction problem in the Schwinger mechanism \cite{tan09}).  Furthermore, higher WKB solutions contain negative powers of $\omega_{\bm p}$.  This leads to severe infrared divergences for massless (or very light) particles, which is actually the case for axion.  }.  Then, it may be natural to assume that the ``true'' mode function should not deviate significantly from that in equilibrium (i.e., the plane wave $\varphi_{\bm p}^{({\rm eq})}$).  The (lowest order) adiabatic particle picture adopts this assumption as a guiding principle to define a mode function in nonequilibrium.  Namely, the adiabatic particle picture approximates the ``true'' mode function by the lowest order Wentzel-Kramers-Brillouin (WKB \cite{wen26, kra26, bri26}) solution $\varphi^{({\rm ad})}_{\bm p}$ of the equation of motion (\ref{eq6b}),
\begin{align}
	\varphi_{\bm p}^{({\rm ad})} \equiv \frac{1}{\sqrt{2\omega_{\bm p}(t)}} \exp \left[  -i\int^{t} dt'\; \omega_{\bm p}(t') \right], \label{eq14}
\end{align}
where
\begin{align}
	\omega_{\bm p}(t) \equiv \sqrt{ {\bm p}^2 + V''[\bar{\phi}(t)] } .
\end{align}
Here, we normalized the adiabatic mode function $\varphi_{\bm p}^{({\rm ad})}$ in the same way as $\varphi_{\bm p}^{({\rm eq})}$ (Eq.~(\ref{eq8})).  Notice that $\omega_{\bm p}$ depends on time $t$ through the time dependence of $\bar{\phi}$ if $V$ is higher than quadratic.  $\varphi_{\bm p}^{({\rm ad})}$ is a natural generalization of $\varphi_{\bm p}^{({\rm eq})}$ as they coincide with each other when the system is in equilibrium, in which $\omega_{\bm p}$ (or $V''$) is merely a constant.  In the adiabatic particle picture, we, then, expand the field operator $\hat{\varphi}$ in terms of the adiabatic mode function $\varphi^{({\rm ad})}_{\bm p}$, and employ the canonical quantization procedure to obtain an adiabatic annihilation operator $\hat{a}^{\rm (ad)}_{\bm p}$ in the same way as in the equilibrium case.  $\hat{a}^{\rm (ad)}_{\bm p}$ satisfies the same commutation relation as $\hat{a}^{\rm (eq)}_{\bm p}$ (Eq.~(\ref{eq10})), and can be deduced from the field operator with $\varphi_{\bm p}^{({\rm ad})}$ as Eq.~(\ref{eq11}).  Nevertheless, there is an important difference between $\hat{a}^{\rm (eq)}_{\bm p}$ and $\hat{a}^{\rm (ad)}_{\bm p}$.  That is, $\hat{a}^{\rm (ad)}_{\bm p}$ depends on time $\hat{a}^{\rm (ad)}_{\bm p} = \hat{a}^{\rm (ad)}_{\bm p}(t)$ because the adiabatic mode function $\varphi^{({\rm ad})}_{\bm p}$ is not an exact solution of the equation of motion (\ref{eq6b}).  This difference results in spontaneous axion particle production as we see below.

\subsubsection{Axion particle production} \label{sec2b2}

We shall show that axion particles are spontaneously produced from the coherent axion field if the axion potential has more than one minimum by explicitly computing the in-vacuum expectation value of the number operator.

To this end, let us for simplicity assume that the system is in equilibrium $V'' = {\rm const.} > 0$ initially $t \leq t_{\rm in}$.  Then, the in-vacuum state $\ket{{\rm vac;}\;t_{\rm in}}$ can be defined as a state such that it is annihilated by $\hat{a}^{\rm (eq)}_{\bm p}$ as
\begin{align}
	0 = \hat{a}^{\rm (eq)}_{\bm p} \ket{{\rm vac;}\;t_{\rm in}}\ \ {\rm for\ any}\ {\bm p}.
\end{align}
On the other hand, by using Eq.~(\ref{eq11}), one can show that the time evolution of the adiabatic annihilation operator $\hat{a}^{\rm (ad)}_{\bm p}$ can be expressed in terms of a Bogoliubov transformation as
\begin{align}
	\hat{a}^{{\rm (ad)}}_{\bm p}(t)
	= \alpha_{\bm p}(t) \hat{a}_{{\bm p}}^{\rm (eq)}  +   \beta_{\bm p}(t) \hat{a}^{{\rm (eq)}\dagger}_{-{\bm p}},  \label{eq15}
\end{align}
where the Bogoliubov coefficients $\alpha_{\bm p}, \beta_{\bm p}$ are given by
\begin{subequations}
\begin{align}
	\alpha_{\bm p} &\equiv i \varphi^{{\rm (ad)}*}_{\bm p} \overset{\leftrightarrow}{\partial}_t  \varphi_{\bm p} = \frac{i \left( \dot{\varphi}_{\bm p} - i\omega_{\bm p} \varphi_{\bm p}  \right)}{\sqrt{2\omega_{\bm p}}} {\rm e}^{+i\int^t \omega_{\bm p} dt}, \\
	\beta_{\bm p}  &\equiv  i \varphi^{{\rm (ad)}*}_{\bm p} \overset{\leftrightarrow}{\partial}_t  \varphi_{-{\bm p}}^{*} = \frac{i \left( \dot{\varphi}^*_{\bm p} - i\omega_{\bm p} \varphi_{\bm p}^*  \right)}{\sqrt{2\omega_{\bm p}}} {\rm e}^{+i\int^t \omega_{\bm p} dt} .
\end{align}
\end{subequations}
Here, $\varphi_{\bm p}$ is the exact solution of the equation of motion (\ref{eq6b}) in the momentum space,
\begin{align}
	0 =  \left[ \partial^2_t + \omega_{\bm p}^2(t) \right] \varphi_{\bm p}, \label{eq16}
\end{align}
with initial conditions
\begin{subequations}
\begin{align}
	\varphi_{\bm p}(t_{\rm in}) &= \varphi_{\bm p}^{\rm (eq)}(t_{\rm in}), \\
	\dot{\varphi}_{\bm p}(t_{\rm in}) &= -i\omega_{\bm p} \varphi_{\bm p}^{\rm (eq)}(t_{\rm in}).
\end{align}
\end{subequations}
Notice that $\beta_{\bm p} = 0$ holds only when $\varphi_{\bm p} = \varphi_{\bm p}^{\rm (ad)}$.  In other words, $\beta_{\bm p} \neq 0$ must hold if the axion potential has more than one minimum.  Indeed, $\varphi_{\bm p}$ couples to the coherent axion field because of the existence of many minima as explained below Eqs.~(\ref{eq6a}) and (\ref{eq6b}), and hence $\varphi_{\bm p}$ should dynamically evolve to deviate from $\varphi_{\bm p}^{\rm (ad)}$ because $\varphi_{\bm p}^{\rm (ad)}$ is no longer a solution of the equation of motion.  Note also that the Bogoliubov coefficients always satisfy
\begin{align}
	1 = |\alpha_{\bm p}|^2 - |\beta_{\bm p}|^2, \label{eq21}
\end{align}
which is a consequence of the unitarity\footnote{A proof is the following: Let us define $S$-matrix as a matrix such that $\hat{a}^{\rm (ad)}_{\bm p}(t) \equiv S^{\dagger} \hat{a}^{\rm (eq)}_{\bm p} S$, and assume $S^\dagger S = {\rm const.} \equiv c > 0$.  Then, we obtain $[ \hat{a}^{\rm (ad)}_{\bm p}(t), \hat{a}^{{\rm (ad)}\dagger}_{\bm p}(t) ] = [ S^{\dagger} \hat{a}^{\rm (eq)}_{\bm p} S, S^{\dagger} \hat{a}^{{\rm (eq)}\dagger}_{\bm p}S] = c^2 \times \delta^3({\bm p}={\bm 0})$.  On the other hand, from Eq.~(\ref{eq15}), one finds $[ \hat{a}^{\rm (ad)}_{\bm p}(t), \hat{a}^{{\rm (ad)}\dagger}_{\bm p}(t) ] = \left( |\alpha_{\bm p}(t)|^2 - |\beta_{\bm p}(t)|^2  \right) \times \delta^3({\bm p}={\bm 0})$.  Therefore, $|\alpha_{\bm p}(t)|^2 - |\beta_{\bm p}(t)|^2 = c^2,$ which says we have Eq.~(\ref{eq21}) only if $S$ is unitary $1 = S^\dagger S$.  }.

It is evident from Eq.~(\ref{eq15}) that the in-vacuum state $\ket{{\rm vac;}\;t_{\rm in}}$ is no longer annihilated by the annihilation operator $\hat{a}_{{\bm p}}^{\rm (ad)}$ for $t > t_{\rm in}$.  Thus, the in-vacuum expectation value of the number operator $\braket{{\rm vac;}\;t_{\rm in}| \hat{a}^{{\rm (ad)}\dagger}_{\bm p} \hat{a}^{{\rm (ad)}}_{\bm p} | {\rm vac;}\;t_{\rm in} }$ is nonvanishing, which implies that axion particles are spontaneously produced from the coherent axion field.  Indeed, one can explicitly compute the expectation value as\footnote{One can derive the same formula on the basis of the nonequilibrium Green function technique \cite{aar01, ber03} and the quantum kinetic theory \cite{smo97, sch98} by assuming quasi-particle picture.  }
\begin{align}
	\frac{d^3N}{d{\bm p}^3}(t)
	&= \braket{{\rm vac;}\;t_{\rm in}|  \hat{a}^{{\rm (ad)}\dagger}_{\bm p}(t) \hat{a}^{{\rm (ad)}}_{\bm p}(t) |{\rm vac;}\;t_{\rm in}}  \nonumber\\
	&= \frac{V}{(2\pi)^3} |\beta_{\bm p}(t)|^2, \label{eq20}
\end{align}
where $\delta^3({\bm p}={\bm 0}) = V/(2\pi)^3$ was used.

\subsection{Energy conservation} \label{sec2c}

Finally, let us discuss the energy balance of the system to check the consistency of our formulation and also the axion particle production from the viewpoint of energy conservation.  To this end, we first explain how to regularize the UV-divergence of two-point functions.  In this paper, we adopt the adiabatic regularization scheme \cite{bir82}, which is compatible with the adiabatic particle picture.  Then, we compute the expectation value of the energy-momentum tensor, and show how the total energy of the system is conserved.  We shall explicitly see that the energy of the coherent axion field is converted into axion particles, and the backreaction term in Eq.~(\ref{eq6a}) is crucial for the energy conservation.

\subsubsection{Adiabatic regularization} \label{sec2C1}

The UV-divergence of in-in expectation values of two-point functions originates from the uninterested vacuum contribution and must be subtracted by a regularization procedure.  Usually, the vacuum is a stable state (i.e., does not decay spontaneously), so that the divergent vacuum contribution is constant in time.  However, in the presence of a time-depending external field, the vacuum state is, in general, no longer stable and it decays spontaneously because of the particle production.  This implies that the vacuum contribution cannot be constant in time (cf. the same problem appears in the real-time dynamics of the Schwinger mechanism \cite{coo89, klu91, klu92, klu93, tan09}).  Indeed, from the time dependence of the adiabatic annihilation operator (\ref{eq15}), one may define the vacuum at time $t$, which we write $\ket{{\rm vac;}\;t}$, as a state such that\footnote{Note that $\ket{{\rm vac;}\;t}$ is a Heisenberg state, so that it is time independent, i.e., $\ket{{\rm vac;}\;t}(t_1) = \ket{{\rm vac;}\;t}(t_2)$ for any $t_1,t_2$.  Nevertheless, as $\ket{{\rm vac;}\;t}$ is defined as an eigen-state of the time-depending adiabatic number operator $\hat{a}^{\rm (ad)}_{\bm p}(t)$ (see Eq.~(\ref{eq-23})), $\ket{{\rm vac;}\;t_1} \neq \ket{{\rm vac;}\;t_2}$ holds if $\hat{a}^{\rm (ad)}_{\bm p}(t_1) \neq \hat{a}^{\rm (ad)}_{\bm p}(t_2)$.  In this sense, one could say $\ket{{\rm vac;}\;t}$ {\it looks} time-dependent although it is a Heisenberg state.  }
\begin{align}
	0 = \hat{a}^{\rm (ad)}_{\bm p}(t) \ket{{\rm vac;}\;t}\ \ {\rm for\ any}\ {\bm p}. \label{eq-23}
\end{align}
Using $\ket{{\rm vac;}\;t}$, one may identify the vacuum contribution at time $t$ of some two-point function $\hat{\varphi} \Gamma \hat{\varphi}$ as
\begin{align}
	\braket{ {\rm vac;}\;t | \hat{ \varphi} \Gamma \hat{\varphi}  |{\rm vac;}\;t }
	= \int \frac{d^3{\bm p}}{(2\pi)^3} \left( \varphi^{\rm (ad)}_{{\bm p}} \Gamma \varphi^{{\rm (ad)}*}_{{\bm p}} \right) ,  \label{eq22}
\end{align}
which obviously depends on time.  Note that it is impossible to express the vacuum contribution at time $t$ (\ref{eq22}) as an expectation value in terms of the in-vacuum $\ket{{\rm vac;}\;{\rm in}}$, which does not know the vacuum structure at time $t$.

In the adiabatic regularization scheme, we subtract the vacuum contribution (\ref{eq22}) from the bare in-vacuum expectation value $\braket{ {\rm vac;}\;t_{\rm in} | \hat{\varphi} \Gamma \hat{\varphi}  |{\rm vac;}\;t_{\rm in} }$, and define the regularized in-vacuum expectation value $\braket{ {\rm vac;}\;t_{\rm in} | : \hat{\varphi} \Gamma \hat{\varphi} : |{\rm vac;}\;t_{\rm in} }$ as
\begin{align}
	&\braket{ {\rm vac;}\;t_{\rm in} | : \hat{\varphi} \Gamma \hat{\varphi} : |{\rm vac;}\;t_{\rm in} } \nonumber\\
		&\equiv \braket{ {\rm vac;}\;t_{\rm in} | \hat{\varphi} \Gamma \hat{\varphi}  |{\rm vac;}\;t_{\rm in} }  -   \braket{ {\rm vac;}\;t | \hat{\varphi} \Gamma \hat{\varphi}  |{\rm vac;}\;t } \nonumber\\
		&= \int \frac{d^3{\bm p}}{(2\pi)^3} \left[  \left( \varphi_{{\bm p}} \Gamma \varphi^{*}_{{\bm p}} \right)  -  \left( \varphi^{({\rm ad})}_{{\bm p}} \Gamma \varphi^{({\rm ad})*}_{{\bm p}} \right) \right].  \label{eq23}
\end{align}
We remark that it is inevitable in the adiabatic regularization scheme to evaluate an expectation value in terms of the adiabatic vacuum state $\ket{{\rm vac;}\;t}$ in order to compute an in-in expectation value in a well-defined/consistent manner.

Let us see explicitly how Eq.~(\ref{eq23}) eliminates the UV-divergence.  For this purpose, let us consider $\Gamma = 1$, i.e., $\hat{\varphi} \Gamma \hat{\varphi} = \hat{\varphi}^2$ as an example.  We first identify the divergent structure of the bare value $\braket{ {\rm vac;}\;t_{\rm in} | \hat{\varphi}^2  |{\rm vac;}\;t_{\rm in} }$ by solving the equation of motion (\ref{eq16}) perturbatively in terms of $1/\omega_{\bm p}$.  To do this, it is convenient to make a WKB-type ansatz for $\varphi_{\bm p}$ as
\begin{align}
	\varphi_{\bm p}(t) = \frac{1}{\sqrt{2\Omega_{\bm p}(t)}} \exp \left[  -i \int^t \Omega_{\bm p}(t')dt' \right].
\end{align}
Then, we substitute this expression into the equation of motion (\ref{eq16}) to find
\begin{align}
	0 = \omega^2_{\bm p} - \Omega^2_{\bm p} + \frac{3}{4}\left(  \frac{\dot{\Omega}_{\bm p}}{\Omega_{\bm p}} \right)^2 - \frac{1}{2} \frac{\ddot{\Omega}_{\bm p}}{\Omega_{\bm p}}.
\end{align}
This equation can be solved perturbatively in terms of $1/\omega_{\bm p}$\footnote{Note that our expansion parameter is the frequency $1/\omega_{\bm p}$.  Thus, the expansion is, precisely speaking, different from the WKB expansion or the adiabatic expansion, whose expansion parameter is $\hbar$ and/or $\omega_{\bm p}^{(n)}$.  } as
\begin{align}
	\Omega_{\bm p}
		&= \omega_{\bm p} + \frac{3\dot{\omega}_{\bm p}^2 - 2\omega_{\bm p} \ddot{\omega}_{\bm p}}{8\omega_{\bm p}^3} + {\mathcal O}(\omega_{\bm p}^{-5}) \nonumber\\
		&= \omega_{\bm p} - \frac{1}{8} \frac{V'''' \dot{\bar{\phi}}^2 + V''' \ddot{\bar{\phi}}}{\omega_{\bm p}^3 }  + {\mathcal O}(\omega_{\bm p}^{-4}).  \label{eq26}
\end{align}
Notice that the first term coincides with the lowest order WKB solution, which is nothing but the adiabatic mode function $\varphi_{\bm p}^{\rm (ad)}$ by definition.  From Eq.~(\ref{eq26}), one can identify the divergent structure of the bare expectation value $\braket{ {\rm vac;}\;t_{\rm in} | \hat{\varphi}^2  |{\rm vac;}\;t_{\rm in} }$ as
\begin{align}
	&\braket{ {\rm vac;}\;t_{\rm in} | \hat{\varphi}^2  |{\rm vac;}\;t_{\rm in} } \nonumber\\
	&= \int \frac{d^3{\bm p}}{(2\pi)^3} \frac{1}{2\Omega_{\bm p}} \nonumber\\
	&= \int \frac{d^3{\bm p}}{(2\pi)^3} \left[   \frac{1}{2\omega_{\bm p}} + \frac{V'''' \dot{\bar{\phi}}^2 + V''' \ddot{\bar{\phi}}}{16\omega_{\bm p}^5} + {\mathcal O}(\omega_{\bm p}^{-6}) \right].
\end{align}
It is evident that only the first term in the brackets is divergent.  In other words, only the contribution from the lowest order WKB solution is UV-divergent, and is always identical to the vacuum contribution\footnote{In general curved spacetime, there can appear additional divergent contributions which cannot be captured by the lowest order WKB solution.  For this case, one has to consider higher order WKB solutions as the adiabatic mode function to define the vacuum state \cite{bir82}.  } 
\begin{align}
	\braket{ {\rm vac;}\;t | \hat{\varphi}^2  |{\rm vac;}\;t }
	&= \int \frac{d^3{\bm p}}{(2\pi)^3} |\varphi_{\bm p}|^2 \nonumber\\
	&= \int \frac{d^3{\bm p}}{(2\pi)^3} \frac{1}{2\omega_{\bm p}}.
\end{align}
Therefore, we have
\begin{align}
	\braket{{\rm in}|:  \hat{\varphi}^2 : |{\rm in}}
	= \int \frac{d^3{\bm p}}{(2\pi)^3} {\mathcal O}(\omega_{\bm p}^{-5}) = ({\rm finite}).  
\end{align}
Thus, the regularized expectation value $\braket{{\rm in}|:  \hat{\varphi}^2 : |{\rm in}} $ defined by the adiabatic regularization scheme (\ref{eq23}) is always finite and well-defined.  Note that $\braket{{\rm in}|:  \hat{\varphi}^2 : |{\rm in}}$ is free from logarithmic divergences.  This is because we dropped higher order terms in $\hbar$ because of the weakness of the axion coupling constant (see Sec.~\ref{sec2a}), i.e., interactions among dynamical axions are neglected (i.e., no loop corrections to mass and/or coupling constants).  If one includes higher order terms in $\hbar$, which are responsible for interactions, a logarithmic divergence would appear (cf. $V = \lambda \phi^4$ case was discussed in Ref.~\cite{coo97}), which can be eliminated by a renormalization procedure \cite{coo89, klu91, klu92, klu93, coo97}.  Also, remark that the adiabatic regularization scheme does not violate the energy conservation law as we will explicitly check below.  The subtraction in nonequilibrium QFT should be time-dependent, so that it is nontrivial whether the subtraction is consistent with conservation laws.  This is not a specific issue in the adiabatic regularization scheme, but is equally nontrivial in other regularization schemes.  Hence, it is important to check conservation laws explicitly so as to ensure that the regularization scheme is not unphysical.

\subsubsection{Energy and pressure}

Now, we evaluate the expectation value of the energy-momentum tensor $T^{\mu\nu}$ within the adiabatic regularization scheme.  The (symmetric) energy-momentum tensor $T^{\mu\nu}$ is defined as
\begin{align}
	\hat{T}_{\mu\nu}
		&\equiv \frac{2}{\sqrt{-g}}\frac{\partial (\sqrt{-g}{\mathcal L})}{\partial g^{\mu\nu}} \nonumber\\
		&= (\partial_{\mu} \hat{\phi})(\partial_{\nu} \hat{\phi}) - g_{\mu\nu} \left[  \frac{1}{2}(\partial_{\lambda} \hat{\phi})(\partial^{\lambda} \hat{\phi}) - V[\hat{\phi}]  \right] \nonumber\\
		&= (\partial_{\mu} \bar{\phi})(\partial_{\nu} \bar{\phi}) - g_{\mu\nu} \left[  \frac{1}{2}(\partial_{\lambda} \bar{\phi})(\partial^{\lambda} \bar{\phi}) - V[\bar{\phi}]  \right] \nonumber\\
		&\quad + 2 (\partial_{\mu} \bar{\phi})(\partial_{\nu} \hat{\varphi}) - g_{\mu\nu} \left[  (\partial_{\lambda} \bar{\phi})(\partial^{\lambda} \hat{\varphi}) - V'[\bar{\phi}]\hat{\varphi}  \right] \nonumber\\
		&\quad + (\partial_{\mu} \hat{\varphi})(\partial_{\nu} \hat{\varphi}) - g_{\mu\nu} \left[ \frac{1}{2} (\partial_{\lambda} \hat{\varphi})(\partial^{\lambda} \hat{\varphi}) - \frac{1}{2}V''[\bar{\phi}] \hat{\varphi}^2  \right],
\end{align}
where higher order quantum corrections ${\mathcal O}(\hbar^{3/2})$ are neglected in the last equality so as to be consistent with Eq.~(\ref{eq5}).

The energy density operator $\hat{\mathcal E}$ and the pressure operator $\hat{P}$ are defined as the diagonal components of $\hat{T}^{\mu\nu}$.  Their expectation value can be evaluated within the adiabatic regularization scheme as
\begin{align}
	&\braket{{\rm vac};\;t_{\rm in}| : \hat{\mathcal E} :| {\rm vac};\;t_{\rm in}} \nonumber\\
	&\equiv \braket{{\rm vac};\;t_{\rm in}| : \hat{T}_{tt} :| {\rm vac};\;t_{\rm in}} \nonumber\\
	&=  \underbrace{ \frac{\left| \partial_t \bar{\phi} \right|^2}{2} + V[\bar{\phi}] }_{\equiv \braket{{\rm vac};\;t_{\rm in}|: \hat{\mathcal E}_{\bar{\phi}} :| {\rm vac};\;t_{\rm in} }} \nonumber\\
	&\quad + \underbrace{ \int \frac{d^3{\bm p}}{(2\pi)^3} \left[ \frac{\left| \partial_t \varphi_{\bm p} \right|^2 + \omega_{\bm p}^2 \left| \varphi_{\bm p} \right|^2  }{2}  - \frac{\omega_{\bm p}}{2} \right] }_{\equiv \braket{{\rm vac};\;t_{\rm in}|: \hat{\mathcal E}_{\varphi}:|{\rm vac};\;t_{\rm in}  } }, \label{eq30}
\end{align}
and
\begin{align}
	&\braket{{\rm vac};\;t_{\rm in}| : \hat{P} :| {\rm vac};\;t_{\rm in}} \nonumber\\
	&\equiv \braket{{\rm vac};\;t_{\rm in}| : \frac{\hat{T}_{xx}+\hat{T}_{yy}+\hat{T}_{zz}}{3} :|{\rm vac};\;t_{\rm in}} \nonumber\\
	&=  \underbrace{ \frac{\left| \partial_t \bar{\phi} \right|^2}{2} - V[\bar{\phi}] }_{\equiv \braket{{\rm vac};\;t_{\rm in}|:\hat{P}_{\bar{\phi}}:| {\rm vac};\;t_{\rm in} }} \nonumber\\
	&\quad + \underbrace{ \int \frac{d^3{\bm p}}{(2\pi)^3} \left[ \frac{\left| \partial_t \varphi_{\bm p} \right|^2 - \left(  \frac{{\bm p}^2}{3} + V''[\bar{\phi}] \right) \left| \varphi_{\bm p} \right|^2  }{2}  - \frac{1}{6}\frac{{\bm p}^2}{\omega_{\bm p}} \right] }_{\equiv \braket{ {\rm vac};\;t_{\rm in}|: \hat{P}_{\varphi} :| {\rm vac};\;t_{\rm in} }}. \label{eq31}
\end{align}
Note that ${\bm \partial} \bar{\phi} = {\bm 0}$ and $\hat{P} = \hat{T}_{xx} = \hat{T}_{yy} = \hat{T}_{zz}$ because we assumed spatial homogeneity.

Let us check the finiteness of the expressions (\ref{eq30}) and (\ref{eq31}).  By using Eq.~(\ref{eq26}), one finds
\begin{align}
	&\braket{{\rm vac};\;t_{\rm in}| : \hat{\mathcal E} :| {\rm vac};\;t_{\rm in}} \nonumber \\
	&= \frac{\left| \partial_t \bar{\phi} \right|^2}{2} + V[\bar{\phi}]  +  \int \frac{d^3{\bm p}}{(2\pi)^3} \left[ \frac{\Omega_{\bm p}}{4} + \frac{\omega_{\bm p}^2}{4\Omega_{\bm p}}+ \frac{\dot{\Omega}_{\bm p}^2}{16\Omega_{\bm p}^3} - \frac{\omega_{\bm p}}{2} \right]  \nonumber\\
	&= \frac{\left| \partial_t \bar{\phi} \right|^2}{2} + V[\bar{\phi}]  +  \int \frac{d^3{\bm p}}{(2\pi)^3} {\mathcal O}(\omega_{\bm p}^{-4}) \nonumber\\
	&= ({\rm finite}),
\end{align} 
and 
\begin{align}
	&\braket{{\rm vac};\;t_{\rm in}| : \hat{P} :| {\rm vac};\;t_{\rm in}} \nonumber\\
	&=  \frac{\left| \partial_t \bar{\phi} \right|^2}{2} - V[\bar{\phi}]   \nonumber\\
		&\quad + \int \frac{d^3{\bm p}}{(2\pi)^3} \left[ \frac{\Omega_{\bm p}}{4} +\frac{\dot{\Omega}_{\bm p}^2}{16\Omega_{\bm p}^3} - \frac{  \frac{{\bm p}^2}{3} + V''[\bar{\phi}] }{4\Omega_{\bm p}} - \frac{{\bm p}^2}{6\omega_{\bm p}} \right] \nonumber\\
	&= \frac{\left| \partial_t \bar{\phi} \right|^2}{2} - V[\bar{\phi}] + \int \frac{d^3{\bm p}}{(2\pi)^3} {\mathcal O}(\omega_{\bm p}^{-4}) \nonumber\\
	&= ({\rm finite}).  
\end{align}
Therefore, the energy density $\braket{{\rm vac};\;t_{\rm in}| : \hat{\mathcal E} :| {\rm vac};\;t_{\rm in}}$ and the pressure $\braket{{\rm vac};\;t_{\rm in}| : \hat{P} :| {\rm vac};\;t_{\rm in}}$ are indeed finite.  Note that there is no logarithmic divergence because there are no interactions among dynamical axions within our order of the $\hbar$-expansion as explained in Sec.~\ref{sec2C1}.

\subsubsection{Energy conservation}

Let us explicitly check the total energy conservation of the system.  The time derivative of the energy density of axion particles can be evaluated as
\begin{align}	
	\partial_t \braket{ {\rm vac};\;t_{\rm in}|: \hat{\mathcal E}_{\varphi} :| {\rm vac};\;t_{\rm in} }= (\partial_t \bar{\phi})  \frac{V'''(\bar{\phi})}{2} \braket{{\rm in}|: \hat{\varphi}^2  :|{\rm in}} , \label{eq32}
\end{align}
where the use is made of the equation of motion for $\varphi_{\bm p}$ (\ref{eq16}).  On the other hand, the time derivative of the energy density of the coherent axion field is given by
\begin{align}
	&\partial_t  \braket{ {\rm vac};\;t_{\rm in}|: \hat{\mathcal E}_{\bar{\phi}} :| {\rm vac};\;t_{\rm in} } \nonumber\\
	&= (\partial_t \bar{\phi}) (\partial^2_t \bar{\phi} + V'(\bar{\phi})) \nonumber\\
	&= - (\partial_t \bar{\phi})  \frac{V'''(\bar{\phi})}{2} \braket{{\rm in}|: \hat{\varphi}^2  :|{\rm in}} \nonumber\\
	&= - \partial_t \braket{ {\rm vac};\;t_{\rm in}|: \hat{\mathcal E}_{\varphi} :| {\rm vac};\;t_{\rm in} } .  \label{eq33}
\end{align}
Here, we used the equation of motion (\ref{eq6a}) to reach the second line.  Therefore, by combining Eqs.~(\ref{eq32}) and (\ref{eq33}), we find that the total energy is strictly conserved as $0 = \partial_t \braket{{\rm vac};\;t_{\rm in}| : \hat{\mathcal E} :| {\rm vac};\;t_{\rm in}}$.

It is now evident from Eq.~(\ref{eq33}) that if the energy of axion particles increases because of the spontaneous axion particle production, the same amount of energy of the coherent axion field must decrease.  That is, the energy of the coherent axion field is the source of axion particle production.  Notice that the right-hand side of Eq.~(\ref{eq33}) vanishes if one neglects the backreaction term (which is higher in $\hbar$) in the equation of motion for the coherent field (\ref{eq6a}).  In this case, axion particles are endlessly created and the energy eventually diverges because the coherent axion field supplies energy to the axion particle production endlessly without any energy loss.  This is clearly unphysical.  Therefore, it is important to maintain the higher order $\hbar^{2}$-term in the equation of motion (\ref{eq6a}) when discussing quantum corrections to the coherent field.

\section{Numerical Results} \label{sec3}

We numerically solve the equations of motion (\ref{eq6a}) and (\ref{eq6b}), and discuss quantitatively the real-time dynamics of the axion particle production from coherent axion field.

\subsection{Setup}

In order to solve the equations of motion (\ref{eq6a}) and (\ref{eq6b}), we need to specify the axion potential $V$ and initial conditions for $\bar{\phi}$ and $\hat{\varphi}$.

For the potential $V$, we consider a periodic potential with infinitely degenerated minima described by
\begin{align}
	V[\phi]  = \chi\left[  1 - \cos\left( \frac{\phi}{f_a} - \theta \right) \right] .  \label{eq34}
\end{align}
Here, $\chi$ is the topological susceptibility of QCD, $f_a$ is the axion decay constant, and $\theta$ is the intrinsic $\theta$-angle.  Notice that one of the stable minima of the potential is $\phi = \theta f_a$, at which
\begin{align}
	\left. V \right|_{\phi = \theta f_a} = \left. V' \right|_{\phi = \theta f_a} = 0, \
	\left. V'' \right|_{\phi = \theta f_a} = \frac{\chi}{f_a^2} \equiv m_0^2
\end{align}
holds.

As an initial condition for coherent axion field $\bar{\phi}$, we impose
\begin{align}
	\left.\bar{\phi}\right|_{t=t_{\rm in}} = f_a \theta, \ \left.\partial_t \bar{\phi} \right|_{t=t_{\rm in}} = \xi.
\end{align}
That is, we consider a situation such that the coherent axion field $\bar{\phi}$ sitting at one of the minima is perturbed suddenly at time $t_{\rm in}$.  Such a perturbation may be seeded by, for example, the misalignment mechanism \cite{abb83, mic83, pre83}\footnote{Precisely, the misalignment mechanism is initiated with finite $\bar{\phi} - f_a \theta \neq 0$ and vanishing $\dot{\bar{\phi}} = 0$.  For this case, the coherent axion field initially has potential energy, instead of kinetic energy.  This difference is essential neither in our production mechanism nor in the evolution of $\bar{\phi}$.  The result is basically the same as long as the initial total energy is the same.  }. QCD sphaleron transitions (e.g. the cosmological QCD phase transition, heavy-ion collisions), and classical electromagnetic fields with nonvanishing ${\bm E}\cdot{\bm B}$ (e.g. solar flares).  The magnitude of $\xi$ determines the initial (kinetic) energy of the coherent field.  In this paper, for the sake of simplicity, $\xi$ is assumed to be a constant which does not depend on ${\bm x}$, i.e., the system is spatially homogeneous.  We also assume $\xi$ is smaller than the QCD energy scale $\xi \lesssim \chi^{1/2} \sim \Lambda_{\rm QCD}^2$ because the axion potential (\ref{eq34}) is an effective potential, which is valid only below the QCD energy scale.

The initial condition for dynamical axion field $\hat{\varphi}$ is already specified in Sec.~\ref{sec2b2}.  We impose that the field operator $\hat{\varphi}$ is expressed in terms of the plane wave at time $t = t_{\rm in}$ as
\begin{align}
	\begin{pmatrix}
		\hat{\varphi}(t_{\rm in},{\bm x}) \\
		\dot{\hat{\varphi}}(t_{\rm in},{\bm x}) 
	\end{pmatrix}
	&= \int d^3{\bm p} \left[ \begin{pmatrix} \varphi^{\rm (eq)}_{\bm p}(t_{\rm in}) \\ \dot{\varphi}^{\rm (eq)}_{\bm p}(t_{\rm in}) \end{pmatrix} \hat{a}^{\rm (eq)}_{\bm p} \right. \nonumber\\
	&\quad\quad\quad\quad \left. +  \begin{pmatrix} \varphi^{{\rm (eq)}*}_{-{\bm p}}(t_{\rm in}) \\ \dot{\varphi}^{{\rm (eq)}*}_{-{\bm p}}(t_{\rm in}) \end{pmatrix} \hat{a}_{-{\bm p}}^{{\rm (eq)}\dagger}  \right] \frac{{\rm e}^{i{\bm p}\cdot{\bm x}}}{(2\pi)^{3/2}},
\end{align}
where
\begin{align}
	\varphi^{\rm (eq)}_{\bm p} = \frac{ 1}{\sqrt{2\sqrt{m_0^2 + {\bm p}^2}}} {\rm e}^{- i \sqrt{m_0^2+{\bm p}^2}t}.
\end{align}

\subsection{Without backreaction} \label{sec3b}

In this subsection, we artificially neglect the backreaction term in the equation of motion for the coherent axion field (\ref{eq6a}).  Although this treatment is unphysical because it violates the energy conservation law as explained in Sec.~\ref{sec2}, it is instructive to understand some basic features of the axion particle production mechanism.  In particular, we numerically show that axion particles are endlessly created, no matter how weak the initial energy $\xi$ is.  This implies that a coherent axion field configuration $\bar{\phi} \neq 0$ is unstable with respect to quantum fluctuations, and thus inclusion of quantum $\hbar$-corrections is inevitable in understanding the real-time dynamics of axion.  Also, an analytical calculation based on the standard perturbation theory is doable in the absence of the backreaction term (see Appendix~\ref{appa}).  We compare the numerical results with the analytical perturbative calculation and see that the present axion particle production mechanism is genuinely nonperturbative which cannot be described within the perturbation theory because it is caused by the nonperturbative quantum tunneling.

\subsubsection{Coherent axion field}

\begin{figure}[t]
 \begin{center}
  \includegraphics[width=0.49\textwidth]{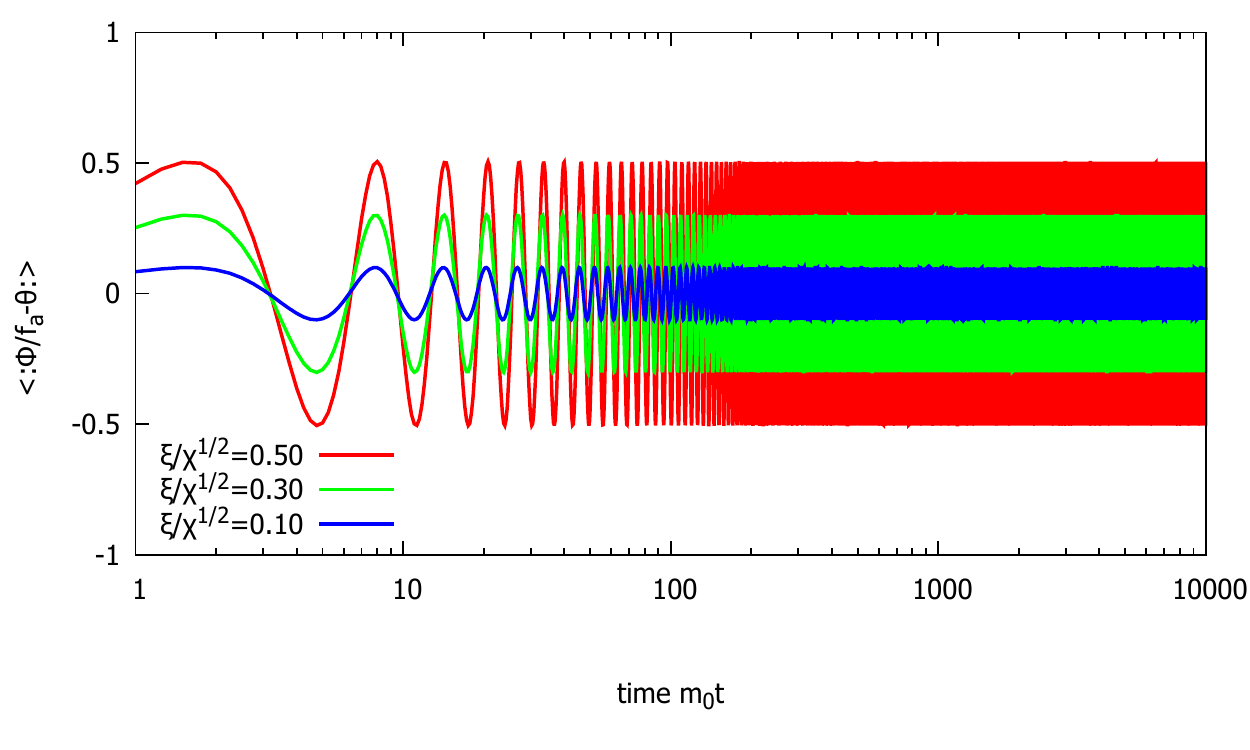}
  \caption{Time evolution of the coherent axion field $\bar{\phi}$ for various value of the initial energy $\xi$ without backreaction. }
  \label{fig1}
 \end{center}
\end{figure}

Figure~\ref{fig1} shows the time evolution of the coherent axion field $\bar{\phi}$.  The coherent field $\bar{\phi}$ oscillates in time with (almost) constant frequency and amplitude.  This is simply because we assumed that the initial kinetic energy of the coherent field is smaller than the potential height $\xi^2 \lesssim \chi$.  Therefore, it is classically forbidden for the coherent field to climb up the potential hill to move to a next minimum, so that it oscillates around the initial minimum just like a harmonic oscillator.  Mathematically, one may expand Eq.~(\ref{eq6a}) in terms of $(\bar{\phi} - \bar{\phi}|_{t=t_{\rm in}})/ \bar{\phi}|_{t=t_{\rm in}} \lesssim 1$ to get
\begin{align}
	0	\sim \left[ \partial_t^2 +  m_0^2 \right] \left( \bar{\phi} - \bar{\phi}|_{t=t_{\rm in}}  \right) . \label{eq39}
\end{align}
One immediately gets a solution to this equation as
\begin{align}
	\bar{\phi}
		&\sim \bar{\phi}|_{t=t_{\rm in}} + \frac{\xi}{m_0} \sin \left(   m_0 t \right).  \label{eq40}
\end{align}
The constant amplitude implies that the coherent field $\bar{\phi}$ never loses energy and endlessly supplies energy to the dynamical field $\hat{\varphi}$ because of the absence of backreaction.  Note that, in expanding systems, there appear additional terms in Eq.~(\ref{eq39}), because of which $\bar{\phi}$ decays in time.

\subsubsection{Variance $\braket{{\rm vac;}\;t_{\rm in}|: \hat{\varphi}^2 :|{\rm vac;}\;t_{\rm in} }$} \label{sec3b2}

\begin{figure}[t]
 \begin{center}
  \includegraphics[width=0.49\textwidth]{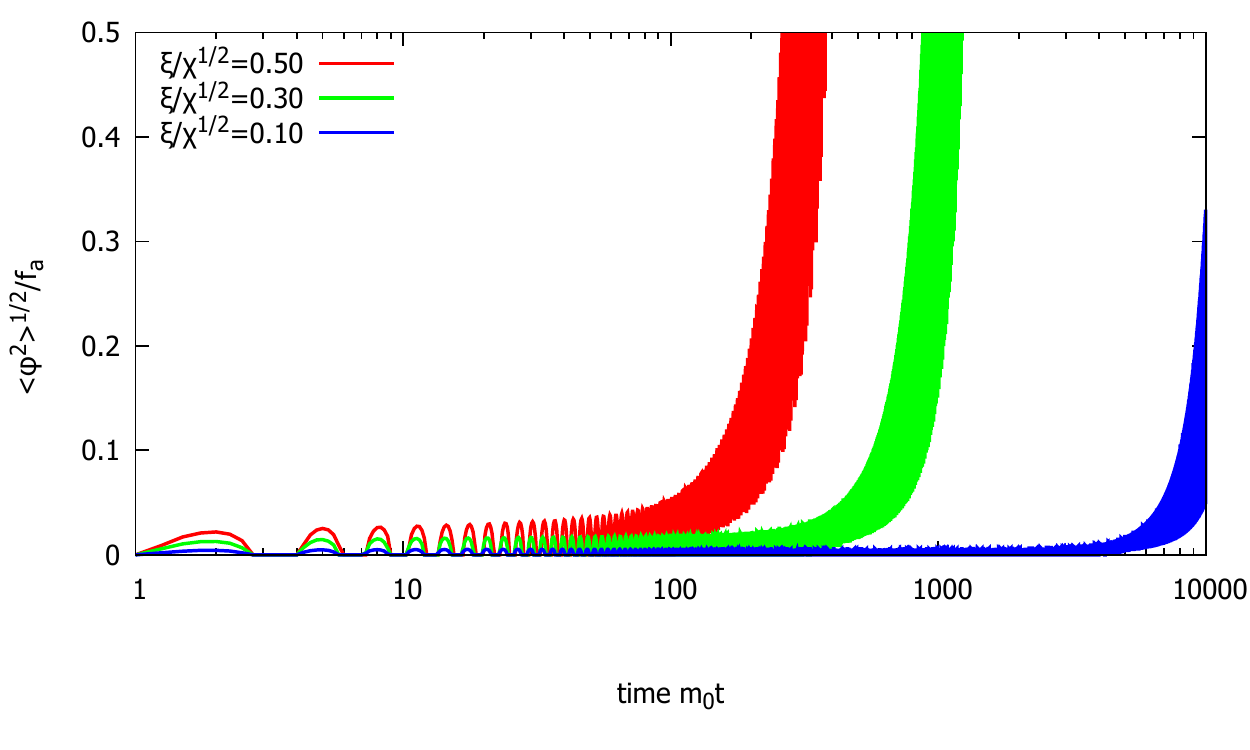}
  \caption{Time evolution of the variance of the axion field $\braket{{\rm vac;}\;t_{\rm in}|: \hat{\varphi}^2 :|{\rm vac;}\;t_{\rm in}} = \braket{{\rm vac;}\;t_{\rm in}|:( \hat{\phi} - \bar{\phi})^2 :|{\rm vac;}\;t_{\rm in}}$ for various values of initial energy $\xi$ without backreaction.  }
  \label{fig2}
 \end{center}
\end{figure}

Figure~\ref{fig2} shows the time evolution of the variance of the axion field (or the magnitude of the dynamical axion field) $\braket{{\rm vac;}\;t_{\rm in}|: \hat{\varphi}^2 :|{\rm vac;}\;t_{\rm in}} = \braket{{\rm vac;}\;t_{\rm in}|: ( \hat{\phi} - \bar{\phi} )^2 :|{\rm vac;}\;t_{\rm in}}$.  The variance $\braket{{\rm vac;}\;t_{\rm in}|: \hat{\varphi}^2 :|{\rm vac;}\;t_{\rm in}}$ diverges exponentially for any values of $\xi$.  That is, although the coherent axion field $\bar{\phi}$ oscillates just around the initial minimum because of the classical restriction, the quantum fluctuation $\hat{\varphi}$ can go far away from the minimum because of quantum tunneling.  The fluctuation eventually extends to infinity because the axion potential (\ref{eq34}) has infinite number of minima, which are distributed infinitely in the field space.  We stress that quantum tunneling plays an essential role here.  Indeed, the fluctuation cannot grow if there exists only one minimum, which is in contrast to the axion particle production via parametric resonance due to oscillating coherent axion field \cite{ray18, sod18, kia18, hay19}.

This quantity $\braket{{\rm vac;}\;t_{\rm in}|: \hat{\varphi}^2 :|{\rm vac;}\;t_{\rm in}}$ is directly related to the total number of produced axions $N/V = \int d^3{\bm p} |\beta_{\bm p}|^2/(2\pi)^3$.  In other words, if $\braket{{\rm vac;}\;t_{\rm in}|: \hat{\varphi}^2 :|{\rm vac;}\;t_{\rm in}}$ increases because of quantum tunneling, $N/V$ should increase, i.e., axion particles are produced.  Indeed, by using
\begin{align}
	\varphi_{\bm p} = \alpha_{\bm p} \varphi_{\bm p}^{({\rm ad})} + \beta^*_{\bm p} \varphi_{\bm p}^{({\rm ad})*},
\end{align}
one can re-express $\braket{{\rm vac;}\;t_{\rm in}|: \hat{\varphi}^2 :|{\rm vac;}\;t_{\rm in}}$ in terms of the Bogoliubov coefficient $\beta_{\bm p}$ as
\begin{align}
	&\braket{{\rm vac;}\;t_{\rm in}|: \hat{\varphi}^2 :|{\rm vac;}\;t_{\rm in}} \nonumber\\
	&= \int \frac{d^3{\bm p}}{(2\pi)^3} \frac{1}{\omega_{\bm p}} \left[ |\beta_{\bm p}|^2 + {\rm Re} \left(  \alpha_{\bm p}\beta_{\bm p} {\rm e}^{-2i\int \omega_{\bm p}dt}   \right)  \right] . \label{eq42}
\end{align}
Note that the second term has an oscillating factor in time, whose frequency is $\sim m_0$.  This is why the curves in Fig.~\ref{fig2} look like a band.  The first term is a classical contribution because $|\beta_{\bm p}|^2$ equals to the phase-space density $(2\pi)^3 d^6N/d{\bm p}^3d{\bm x}^3$ (\ref{eq20}).  The second term arises because of a quantum interference between the positive and negative frequency modes, and gives a quantum correction to the classical value.  Usually, the second term gives only a small contribution compared to the first one after the momentum integration because the integrand contains severely oscillating factor in momentum $\sim \exp \left[-i \int \omega_{\bm p}dt \right]$.  As we shall see soon, produced axions are typically soft $|{\bm p}| \lesssim m_0$ (see Fig.~\ref{fig4}), for which energy cost to produce becomes smaller.  Thus, it is good to approximate $\omega_{\bm p}$ in Eq.~(\ref{eq42}) as $\omega_{\bm p} \sim m_0$.  Then, after neglecting the quantum correction, one obtains
\begin{align}
	\braket{{\rm vac;}\;t_{\rm in}|: \hat{\varphi}^2 :|{\rm vac;}\;t_{\rm in}}
		\sim \frac{1}{m_0} \frac{N}{V}.  \label{eq43}
\end{align}

\subsubsection{Axion particle production} \label{sec3b3}

\begin{figure}[t]
 \begin{center}
  \includegraphics[width=0.49\textwidth]{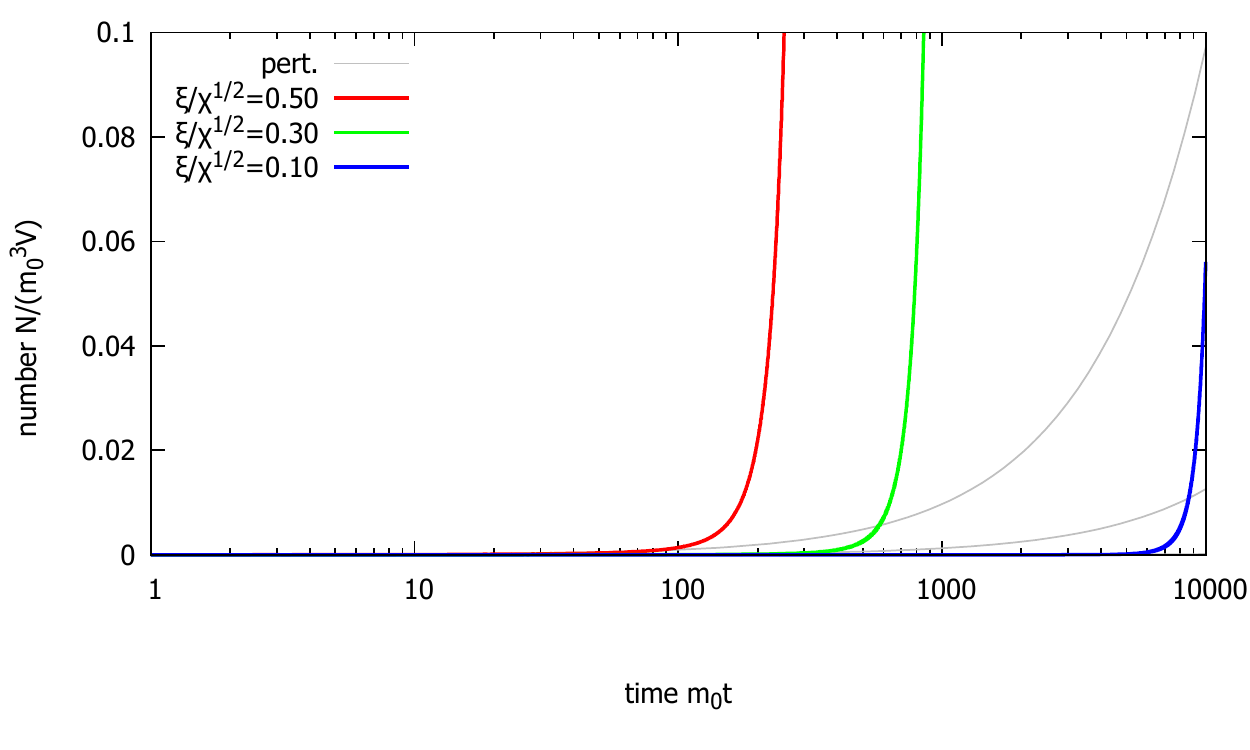}
  \caption{Time evolution of the number density $N/V$ of axion particles for various values of initial energy $\xi$ without backreaction.  }
  \label{fig3}
 \end{center}
\end{figure}
\begin{figure}[t]
 \begin{center}
  \includegraphics[width=0.49\textwidth]{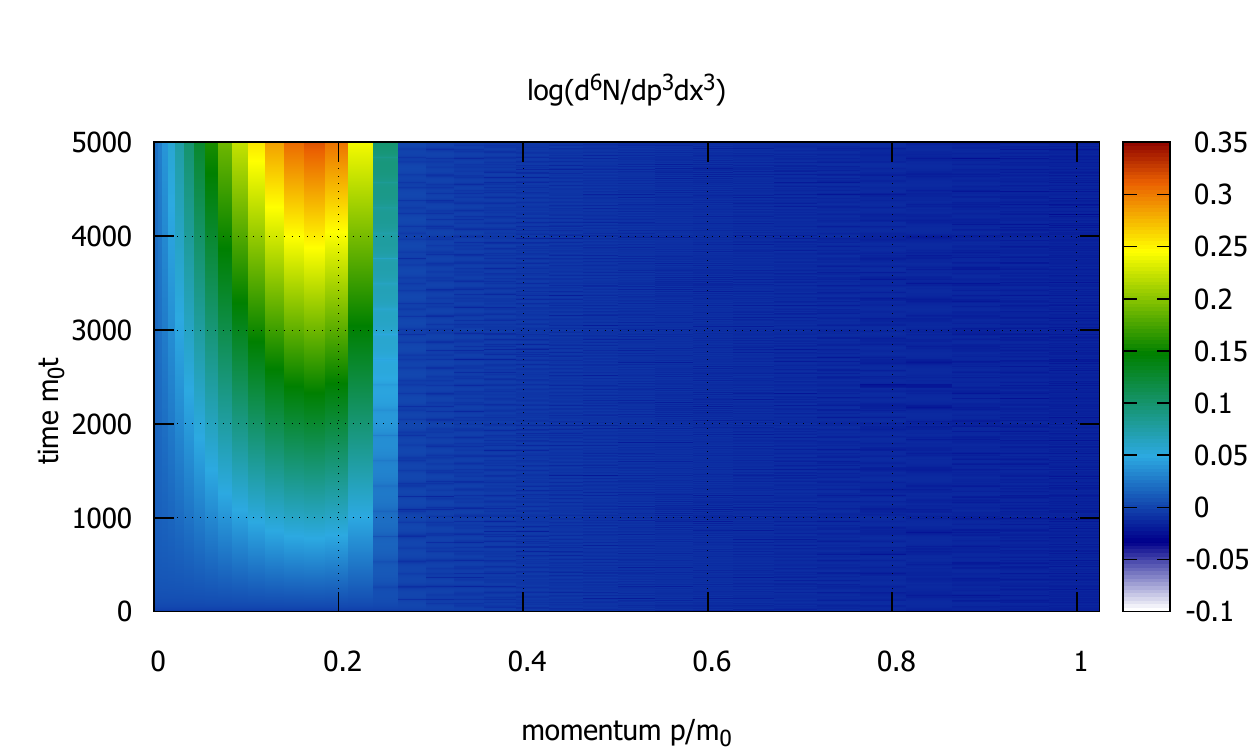}
  \caption{Time evolution of the phase-space density $(2\pi)^3 d^6N/d{\bm p}^3d{\bm x}^3 = |\beta_{\bm p}|^2$ of axion particles without backreaction.  The initial energy $\xi$ is fixed as $\xi/\sqrt{\chi}=0.5$.  }
  \label{fig4}
 \end{center}
\end{figure}

Figure~\ref{fig3} shows the time evolution of the total number of produced axion particles $N/V = \int d^3{\bm p} |\beta_{\bm p}|^2/(2\pi)^3 $.  $N/V$ increases exponentially for any value of $\xi$.  This implies that the coherent axion field is unstable against quantum fluctuations and it spontaneously decays by producing axion particles no matter how small or large the initial fluctuation $\xi$ is.

The time evolution of the phase-space density $(2\pi)^3 d^6N/d{\bm p}^3d{\bm x}^3$ of produced axion particles is shown in Fig.~\ref{fig4}.  Low momentum $|{\bm p}|\lesssim m_0$ axion particles dominate the production.  This is a natural result because it is energetically costful to create hard particles.

It is instructive to compare these results with a perturbative calculation to clarify the nonperturbative nature of our production mechanism: if the initial energy is sufficiently small compared to the QCD energy scale $\xi \lesssim \sqrt{\chi}$, the coherent axion field does not deviate significantly from the initial minimum $\delta \bar{\phi} \equiv (\bar{\phi} - \bar{\phi}_{t=t_{\rm in}})/\bar{\phi}_{t=t_{\rm in}} \lesssim 1$ (see Fig.~\ref{fig1}).  Then, one may naively expect that the axion particle production rate can be evaluated perturbatively in terms of $\delta \bar{\phi}$ (or equivalently $\xi$).  As the coherent field $\bar{\phi}$ decouples from the dynamical field $\hat{\varphi}$ in the absence of the backreaction term, one can carry out the perturbative calculation analytically as (see Appendix~\ref{appa} for the detail)
\begin{align}
	\frac{d^7N^{\rm (pert)}}{d{\bm p}^3d{\bm x}^3dt}
	= \frac{m_0^4}{(64\pi)^2} \left(\frac{\xi}{\sqrt{\chi} }\right)^4 \frac{\delta(|{\bm p}|)}{{\bm p}^2} + {\mathcal O}\left(  \left(\frac{\xi}{\sqrt{\chi} }\right)^6  \right),
\end{align}
and
\begin{align}
	\frac{1}{V}\frac{dN^{\rm (pert)}}{dt} = \frac{m_0^4}{2048\pi} \left(\frac{\xi}{\sqrt{\chi} }\right)^4  + {\mathcal O}\left(  \left(\frac{\xi}{\sqrt{\chi} }\right)^6  \right).
\end{align}
An important point here is that the perturbative production number $N^{\rm (pert)}$ is proportional to $t$.  Apparently, this is inconsistent with the exponential growth observed in Figs.~\ref{fig3} and \ref{fig4}, and the perturbation theory agrees with our results only at very early times.  This disagreement is a natural result because our production mechanism is a consequence of quantum tunneling (as we discussed in Sec.~\ref{sec3b2}), in which the existence of many minima is important.  By the perturbative expansion in terms of $\delta \bar{\phi}$, only the local structure of the axion potential around one minimum is taken into account, and the global information of the axion potential (i.e., the existence of many minima) is lost.

\subsubsection{Energy and pressure} \label{sec314}

\begin{figure}[t]
 \begin{center}
  \includegraphics[width=0.49\textwidth]{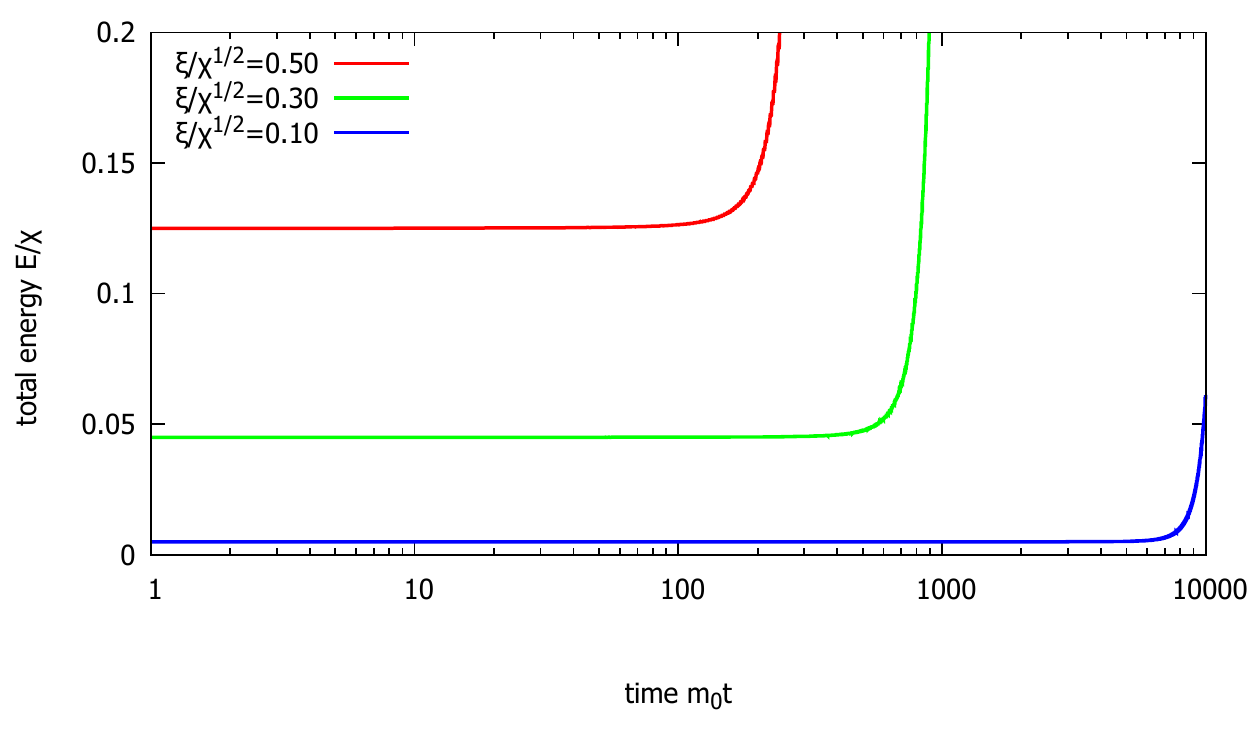}
  \includegraphics[width=0.49\textwidth]{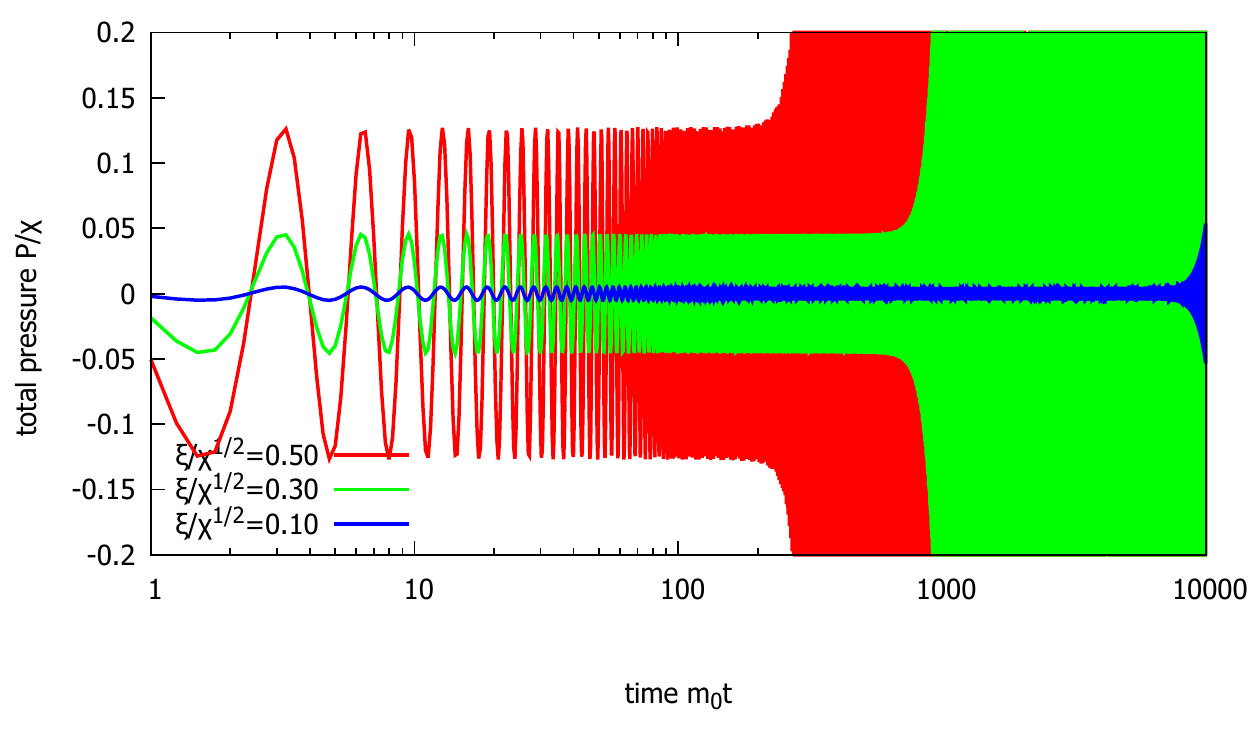}
  \caption{Time evolution of the total energy density $\braket{{\rm vac};t_{\rm in}|: \hat{\mathcal E} :|{\rm vac};t_{\rm in}}$ (top) and the total pressure $\braket{{\rm vac};t_{\rm in}|: \hat{P} :|{\rm vac};t_{\rm in}}$ (bottom) for various values of initial energy $\xi$ without backreaction.  }
  \label{fig5}
 \end{center}
\end{figure}

Figure~\ref{fig5} shows the time evolution of the total energy density $\braket{{\rm vac;}\;t_{\rm in}|: \hat{\mathcal E} :|{\rm vac;}\;t_{\rm in}}$ and total pressure $\braket{{\rm vac;}\;t_{\rm in}|: \hat{P} :|{\rm vac;}\;t_{\rm in}}$.  As expected, the energy and the pressure diverge because of the divergent contribution from axion particles $\hat{\varphi}$.  Therefore, one cannot dismiss the backreaction term to describe real-time dynamics of axion.

The energy density is dominated by the mass energy of produced axion particles because they are soft (see Fig.~\ref{fig4}).  To see this analytically and to find out relationship to other observables such as $N$ and the variance, let us rewrite $\braket{{\rm vac;}\;t_{\rm in}|: \hat{\mathcal E}_{\varphi} :|{\rm vac;}\;t_{\rm in}}$ in terms of the Bogoliubov coefficient.  By using Eq.~(\ref{eq30}), one obtains
\begin{align}
	\braket{{\rm vac;}\;t_{\rm in}|: \hat{\mathcal E}_{\varphi} :|{\rm vac;}\;t_{\rm in}}
		= \int \frac{d^3{\bm p}}{(2\pi)^3} \omega_{\bm p} |\beta_{\bm p}|^2 .
\end{align}
Then, by approximating $\omega_{\bm p}\sim m_0$, we find
\begin{align}
	\braket{{\rm vac;}\;t_{\rm in}|: \hat{\mathcal E}_{\varphi} :|{\rm vac;}\;t_{\rm in}} 
		&\sim m_0 \times \frac{N}{V} \nonumber\\
		&\sim m_0^2 \braket{{\rm in}|: \hat{\varphi}^2 :|{\rm in}}.
\end{align}

On the other hand, quantum corrections dominate the pressure.  To show this, let us rewrite $\braket{{\rm vac;}\;t_{\rm in}|: \hat{P}_{\varphi} :|{\rm vac;}\;t_{\rm in}}$ (see Eq.~(\ref{eq31})) in terms of the Bogoliubov coefficient as
\begin{align}
	&\braket{{\rm vac;}\;t_{\rm in}|: \hat{P}_{\varphi} :|{\rm vac;}\;t_{\rm in}} \nonumber\\ 
	&= \int \frac{d^3{\bm p}}{(2\pi)^3} \left[ |\beta_{\bm p}|^2 \frac{{\bm p}^2/3}{\omega_{\bm p}}  +  \frac{{\bm p}^2/3 - \omega_{\bm p}^2}{\omega_{\bm p}} {\rm Re}\left[ \alpha_{\bm p}\beta_{\bm p} {\rm e}^{-2i\int \omega_{\bm p} dt} \right] \right].  \label{eq48}
\end{align}
As we know that produced axion particles are soft (see Fig.~\ref{fig4}), it is good to approximate $\omega_{\bm p} \sim m_0$.  Then, only the second term in Eq.~(\ref{eq48}), which is nothing but quantum corrections, survives as
\begin{align}
	\braket{{\rm in}|: \hat{P}_{\varphi} :|{\rm in}}
	\sim - \int \frac{d^3{\bm p}}{(2\pi)^3} {\rm Re}\left[   \alpha_{\bm p}\beta_{\bm p} {\rm e}^{-2i\int \omega_{\bm p} dt} \right] .  \label{eq49}
\end{align}
Equation~(\ref{eq49}) oscillates in time, which is the reason why we see an oscillating behavior in Fig.~\ref{fig5}.

\subsection{With backreaction}

Let us recover the backreaction term, and study the role of the energy conservation.  Below, we explicitly see that because of the energy conservation, (i) dynamics of the coherent field $\bar{\phi}$ is no longer independent of the dynamical field $\hat{\varphi}$; (ii) the variance cannot extend to infinity, i.e., $\braket{{\rm vac};\;t_{\rm in}|: \hat{\varphi}^2 :|{\rm vac};\;t_{\rm in}} < \infty $; (iii) the number of produced axion particles stays finite; and (iv) not only the axion particle production process, but also an axion particle pair annihilation process takes place.

\subsubsection{Coherent axion field} \label{sec411}

\begin{figure}[t]
 \begin{center}
  \includegraphics[width=0.49\textwidth]{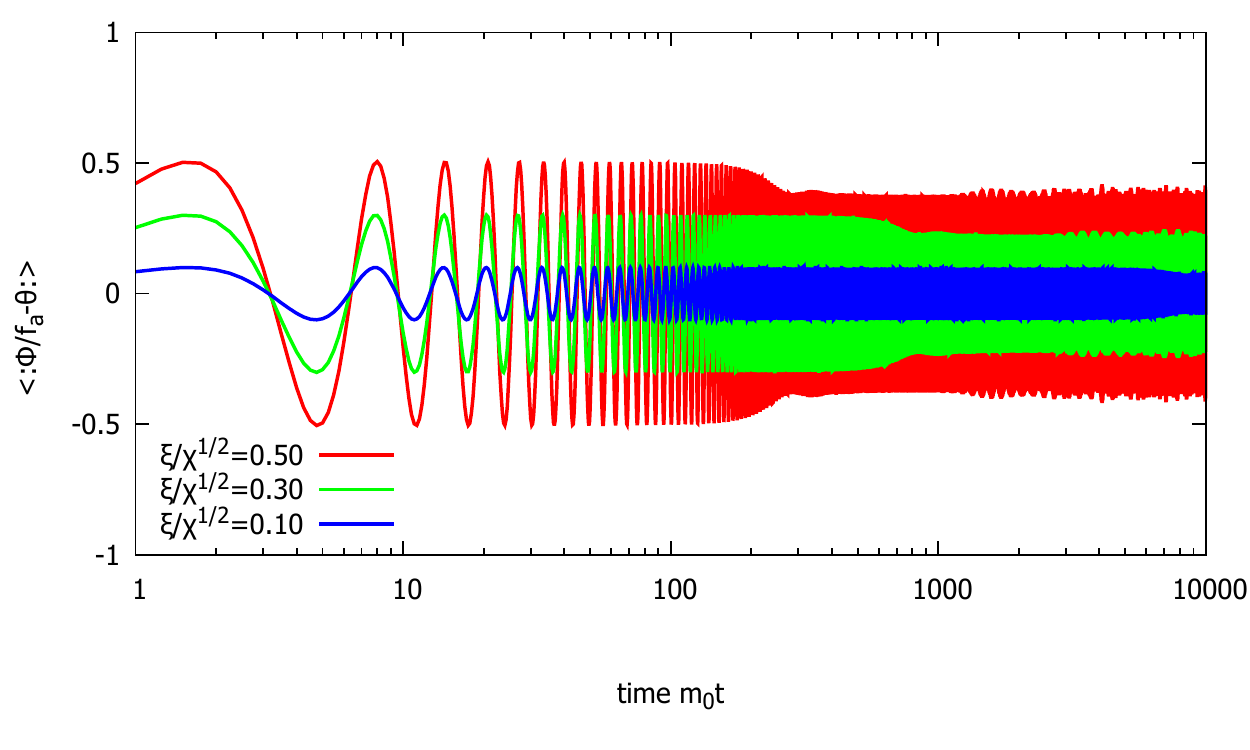}
  \caption{Time evolution of the coherent axion field $\bar{\phi}$ for various value of the initial energy $\xi$ with backreaction.  }
  \label{fig6}
 \end{center}
\end{figure}

Figure~\ref{fig6} shows the time evolution of the coherent field $\bar{\phi}$.  Since the energy of the coherent field is consumed in the axion particle production, the amplitude of $\bar{\phi}$ decreases from the initial value.  However, the amplitude never goes to zero even at late times.  In other words, the coherent field never falls into a minimum of the potential (i.e., no equilibration), but it oscillates around a minimum forever if the system is non-expanding.

This is because axion particle pair annihilation occurs.  Indeed, because of the backreaction term, not only the axion particle production process, but also its inverse process, i.e., pair annihilation process ($2\hat{\varphi} \to \bar{\phi}$) can occur.  Due to the pair annihilation process, the time evolution of the coherent field is modified as follows: For the first moments, the coherent field loses its energy (i.e., amplitude decreases) and the phase-space density of axion particles increases because of the axion particle production process.  As the phase-space density increases, produced axion particles begin to overlap with each other and the pair annihilation process becomes more efficient than the production process.  Once the pair annihilation process dominates, the phase-space density decreases and the coherent field would obtain energy from axion particles (i.e., amplitude increases) because of the energy conservation; see Eq.~(\ref{eq33}).  After some time, the phase-space density becomes dilute and the pair annihilation process should be suppressed.  Then, the production process would again be more efficient.  In this way, the dominant process changes as time goes, and hence the coherent field never gets equilibrated.  Notice that we can see small oscillations in the amplitude at late times in Fig.~\ref{fig6}.  This is a manifestation of the above mechanism.  Also, in Sec.~\ref{sec413}, we shall discuss the time evolution of the phase-space density.  There, we can explicitly see that the annihilation process and the production process, indeed, alternatively take place.  Note that the same oscillation mechanism was already discussed in the real-time dynamics of the Schwinger mechanism \cite{klu92, klu93, tan09}, in which classical electric field shows an oscillating behavior (plasma oscillation).  Note also that in order for the coherent field to completely fall into a minimum, the energy of the total axion field $\hat{\phi}$ must be dissipated by, for example, coupling to other particles and attaching to heat bath.

The frequency of the oscillation deviates slightly from $m_0$ because of the backreaction term.  Indeed, by assuming that $\braket{{\rm vac};\;t_{\rm in}|: \hat{\varphi}^2 :|{\rm vac};\;t_{\rm in}}$ is sufficiently adiabatic, one can expand the equation of motion (\ref{eq6a}) in terms of $(\bar{\phi} - \bar{\phi}|_{t=t_{\rm in}})/ \bar{\phi}|_{t=t_{\rm in}} \lesssim 1$ as
\begin{align}
	0 &\sim \left[ \partial_t^2 +  \left( m_0 \sqrt{1-\frac{\braket{{\rm vac};\;t_{\rm in}|: \hat{\varphi}^2 :|{\rm vac};\;t_{\rm in}}}{2f_a^2}} \right)^2 \right] \nonumber\\
	&\quad \times \left( \bar{\phi} - \bar{\phi}|_{t=t_{\rm in}}  \right) .
\end{align}
Thus, the frequency is given, roughly, by $m_0 \times \sqrt{1-\braket{{\rm vac};\;t_{\rm in}|: \hat{\varphi}^2 :|{\rm vac};\;t_{\rm in}}/2f_a^2} < m_0$.  Intuitively, produced axion particles act as friction to the coherent field, which lowers the frequency of the oscillation.

\subsubsection{Variance $\braket{{\rm vac};t_{\rm in}| \hat{\varphi}^2 |{\rm vac};t_{\rm in}}$}

\begin{figure}[t]
 \begin{center}
  \includegraphics[width=0.49\textwidth]{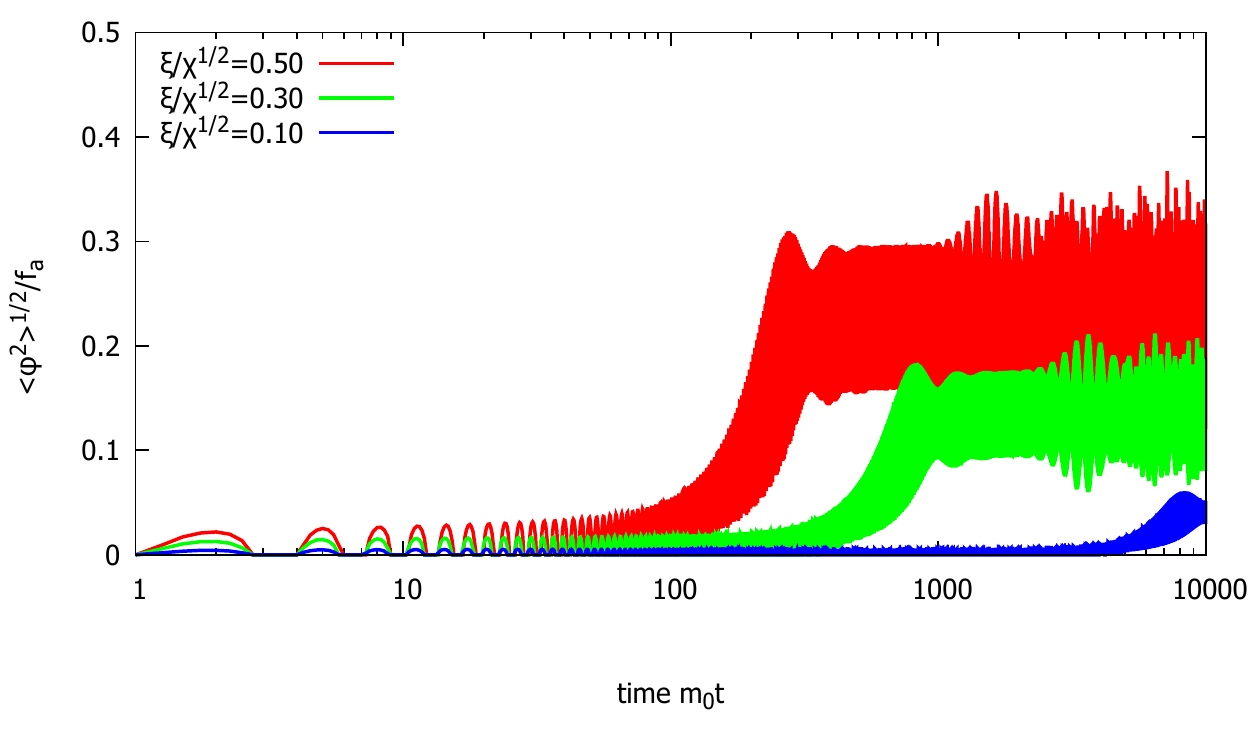}
  \caption{Time evolution of the variance of the axion field $\braket{{\rm vac;}\;t_{\rm in}|: \hat{\varphi}^2 :|{\rm vac;}\;t_{\rm in}} = \braket{{\rm vac;}\;t_{\rm in}|:( \hat{\phi} - \bar{\phi} )^2 :|{\rm vac;}\;t_{\rm in}}$ for various values of initial energy $\xi$ with backreaction.  }
  \label{fig7}
 \end{center}
\end{figure}

Figure~\ref{fig7} shows the time evolution of the variance $\braket{{\rm vac};\;t_{\rm in}|: \hat{\varphi}^2 :|{\rm vac};\;t_{\rm in}}$.  For the first moments, where the number of produced axion particles is small and the backreaction effect can be neglected, the variance grows exponentially as we found in Fig.~\ref{fig2}.  However, after sufficient number of axion particles is produced, the backreaction effect becomes important and the variance stops growing.  Thus, as expected, the variance never diverges, i.e., the quantum fluctuation $\hat{\varphi}$ cannot go far away from the initial minimum because of the energy conservation.  Note that the variance shows oscillating behaviors at late times because of the pair annihilation process as explained in Sec.~\ref{sec411}.

\subsubsection{Axion particle production} \label{sec413}

\begin{figure}[t]
 \begin{center}
  \includegraphics[width=0.49\textwidth]{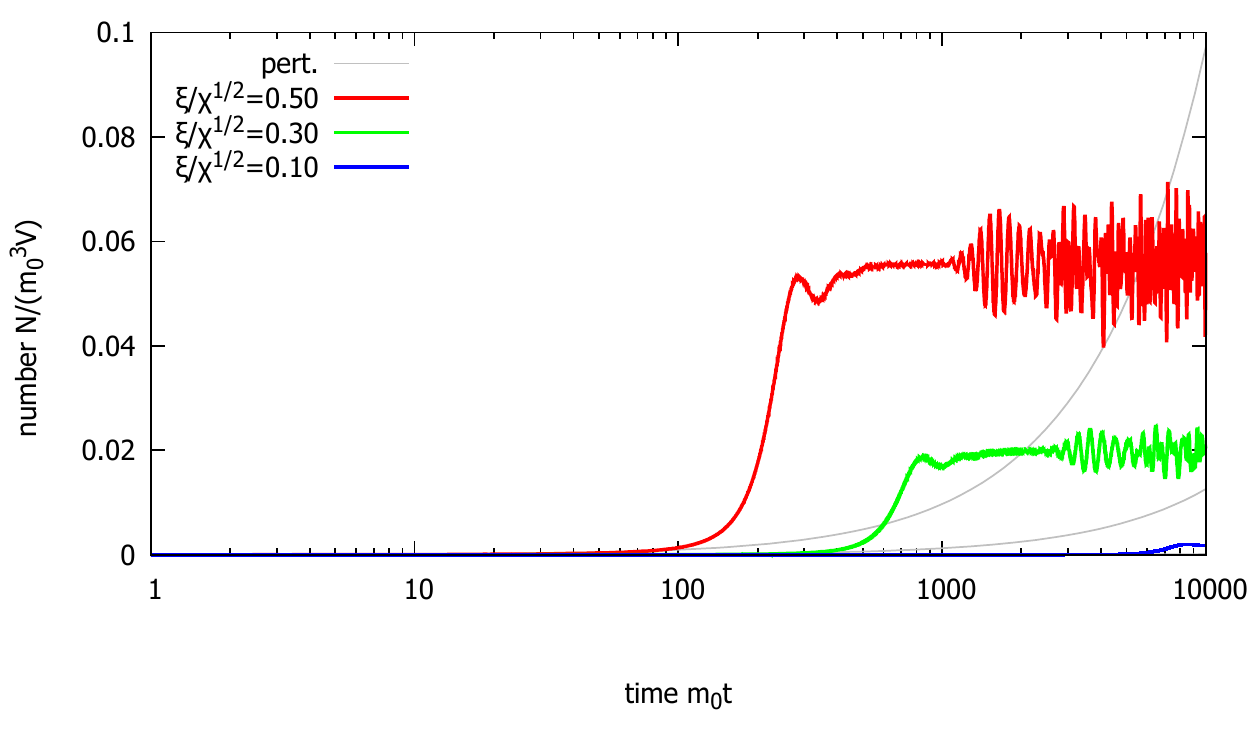}
  \caption{Time evolution of the number density $N/V$ of axion particles for various values of initial energy $\xi$ with backreaction.  }
  \label{fig8}
 \end{center}
\end{figure}

\begin{figure}[t]
 \begin{center}
  \includegraphics[width=0.49\textwidth]{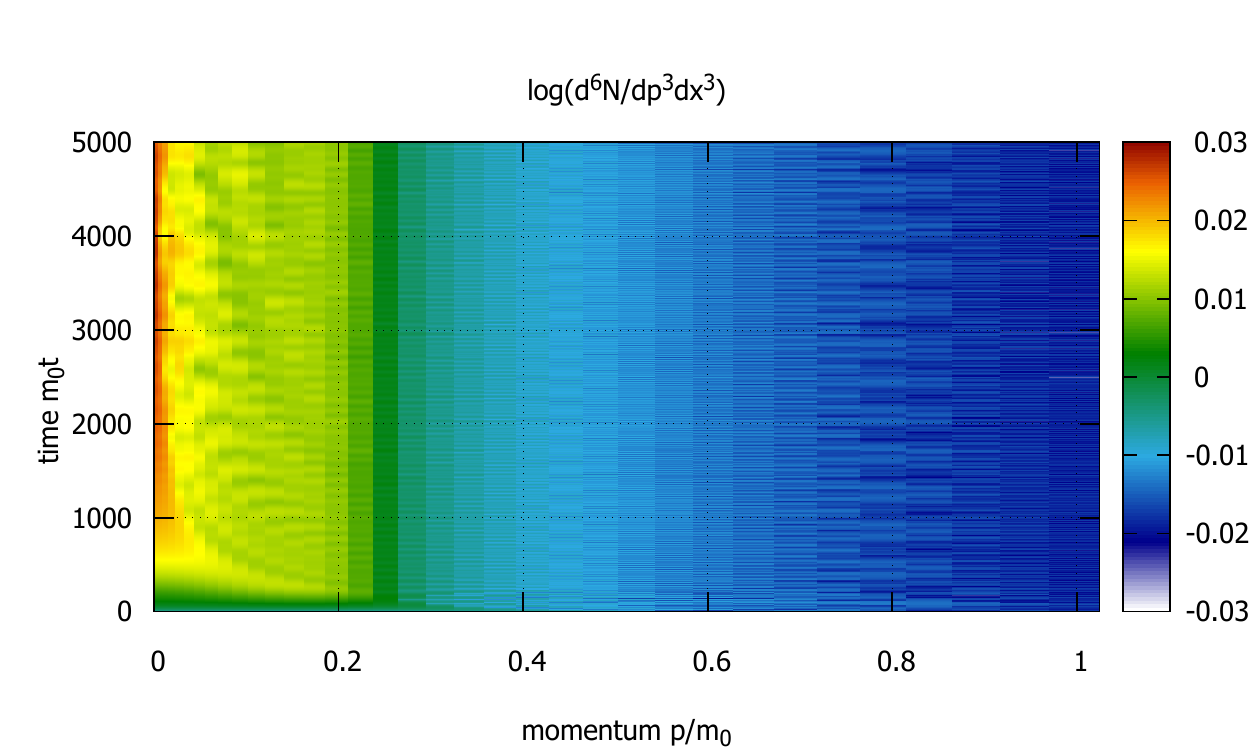}
  \caption{Time evolution of the phase-space density $(2\pi)^3 d^6N/d{\bm p}^3d{\bm x}^3 = |\beta_{\bm p}|^2$ of axion particles with backreaction.  The initial energy $\xi$ is fixed as $\xi/\sqrt{\chi}=0.5$.   }
  \label{fig9}
 \end{center}
\end{figure}

Figure~\ref{fig8} shows the total number $N/V = \int d^3{\bm p} |\beta_{\bm p}|^2/(2\pi)^3 $ of produced axion particles.  It is evident that the total number stays finite because of the energy conservation.  The time dependence is basically the same as the variance $\braket{{\rm vac};\;t_{\rm in}|: \hat{\varphi}^2 :|{\rm vac};\;t_{\rm in}}$ (see Fig.~\ref{fig7}) because they are closely related with each other as shown by Eq.~(\ref{eq43}).  The only difference is that we cannot see the ``band structure'' in this figure because the variance receives quantum corrections (the second term in Eq.~(\ref{eq42})), which oscillate severely in time.

Figure~\ref{fig9} shows the time evolution of the phase-space density  $(2\pi)^3d^6N/d{\bm p}^3d{\bm x}^3$ of produced axion particles.  The phase-space density is significantly modified by the backreaction (see Fig.~\ref{fig4}).  In particular, the pair annihilation process plays an important role: The phase-space density is softened by pair annihilation of hard axion particles.  Also, we can see a striped pattern in the figure, which is because the annihilation process and the production process alternatively take place as explained in Sec.~\ref{sec411}.

With our parameter choice $\xi/\sqrt{\chi} = {\mathcal O}(0.1)$, i.e., the initial energy of the coherent axion field is at most the QCD energy scale, the axion particle production is completed, roughly, within $t_{\rm prod} \sim {\mathcal O}(1000) \times m_0^{-1}$.  The current axion mass bound is $10^{-5}\;{\rm eV} \lesssim m_0 \lesssim 10^{-2}\;{\rm eV}$, so that the time-scale can be estimated as $10^{5}\;{\rm eV}^{-1} \lesssim t_{\rm prod} \lesssim 10^{8}\;{\rm eV}^{-1}$ (or $10^{-10}\;{\rm sec} \lesssim t_{\rm prod} \lesssim 10^{-7}\;{\rm sec}$).

\subsubsection{Energy and pressure}

\begin{figure*}[t]
 \begin{center}
  \hspace{-5.5mm}
  \includegraphics[width=0.355\textwidth]{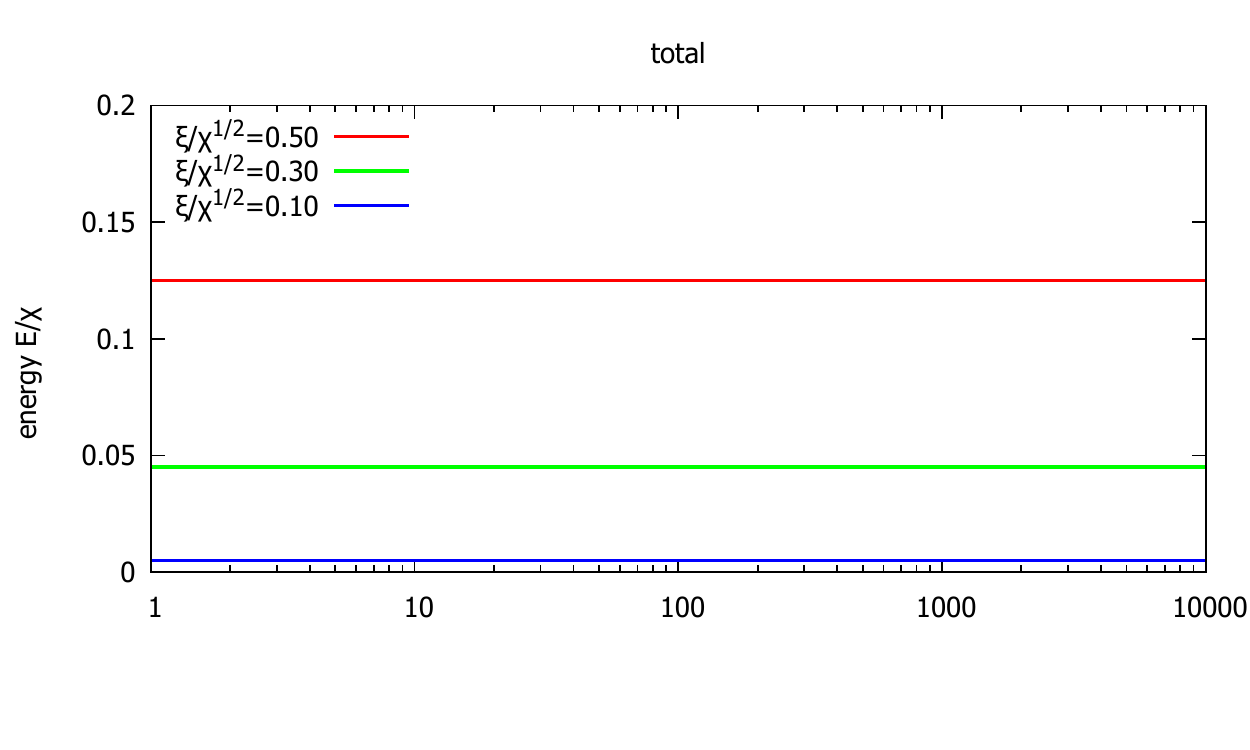} \hspace{-5.5mm}
  \includegraphics[width=0.355\textwidth]{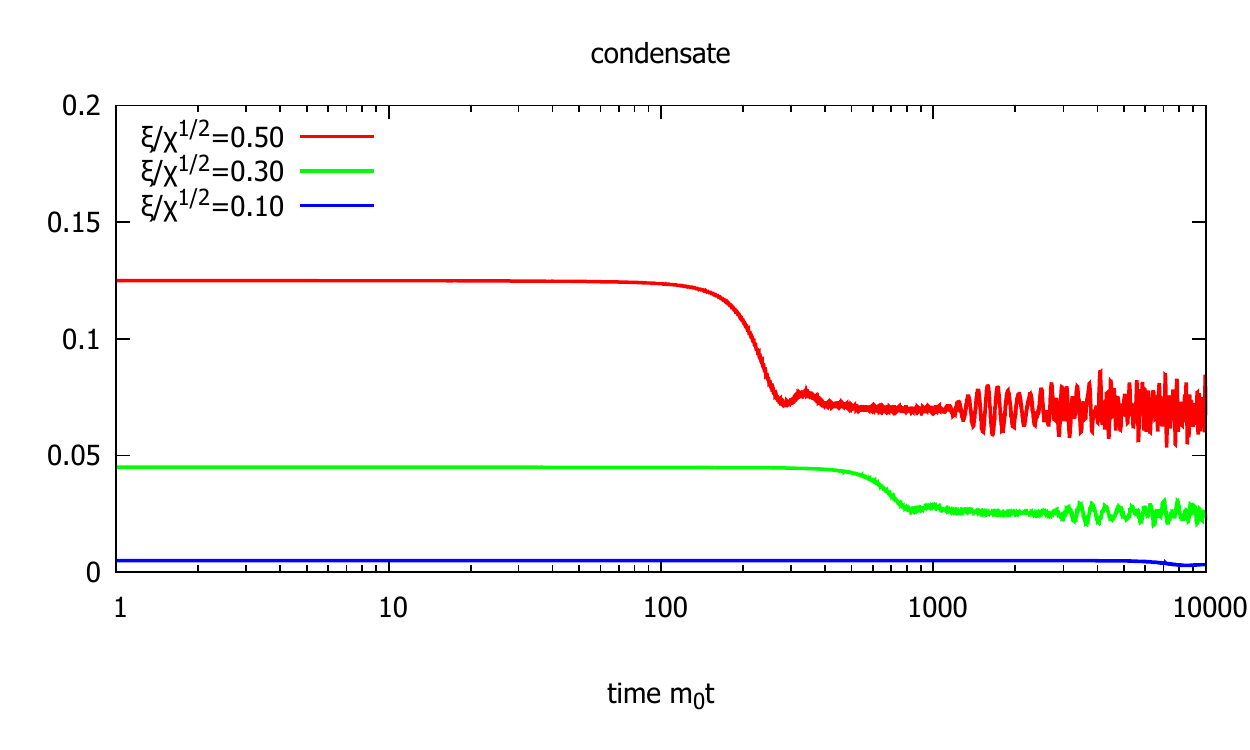} \hspace{-5.5mm}
  \includegraphics[width=0.355\textwidth]{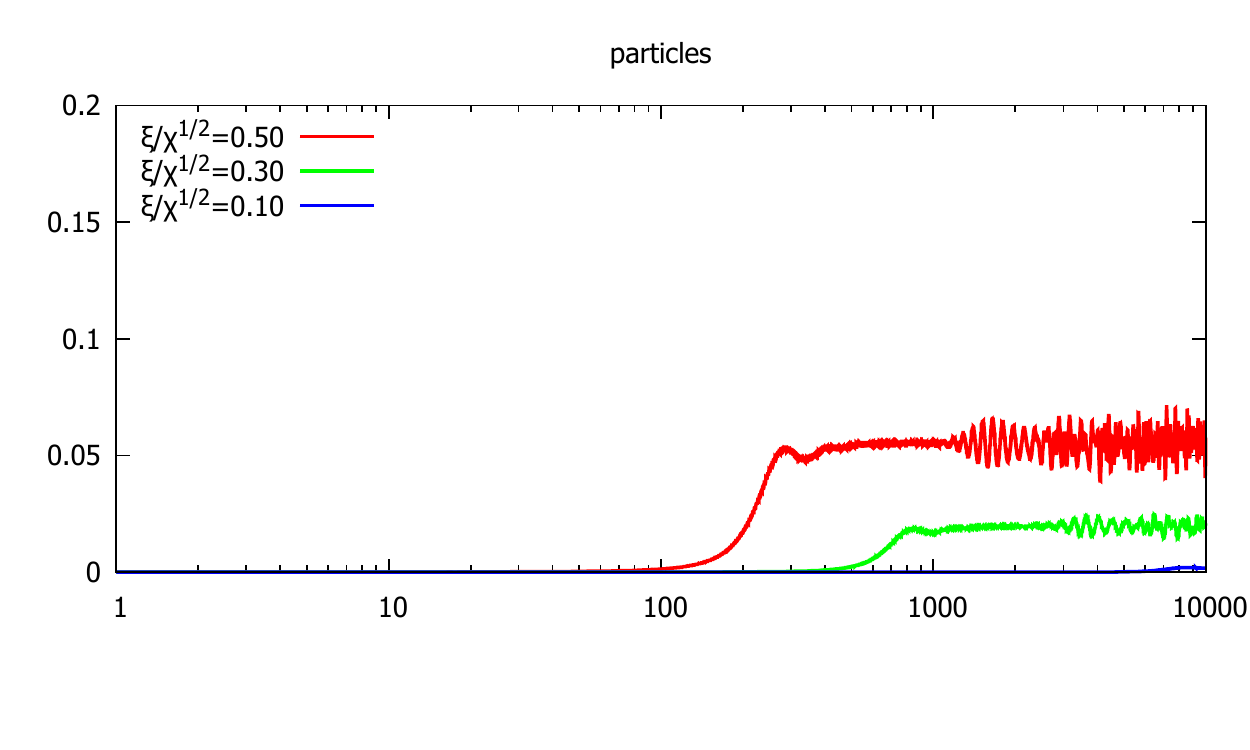} \\
  \hspace{-5.5mm}
  \includegraphics[width=0.355\textwidth]{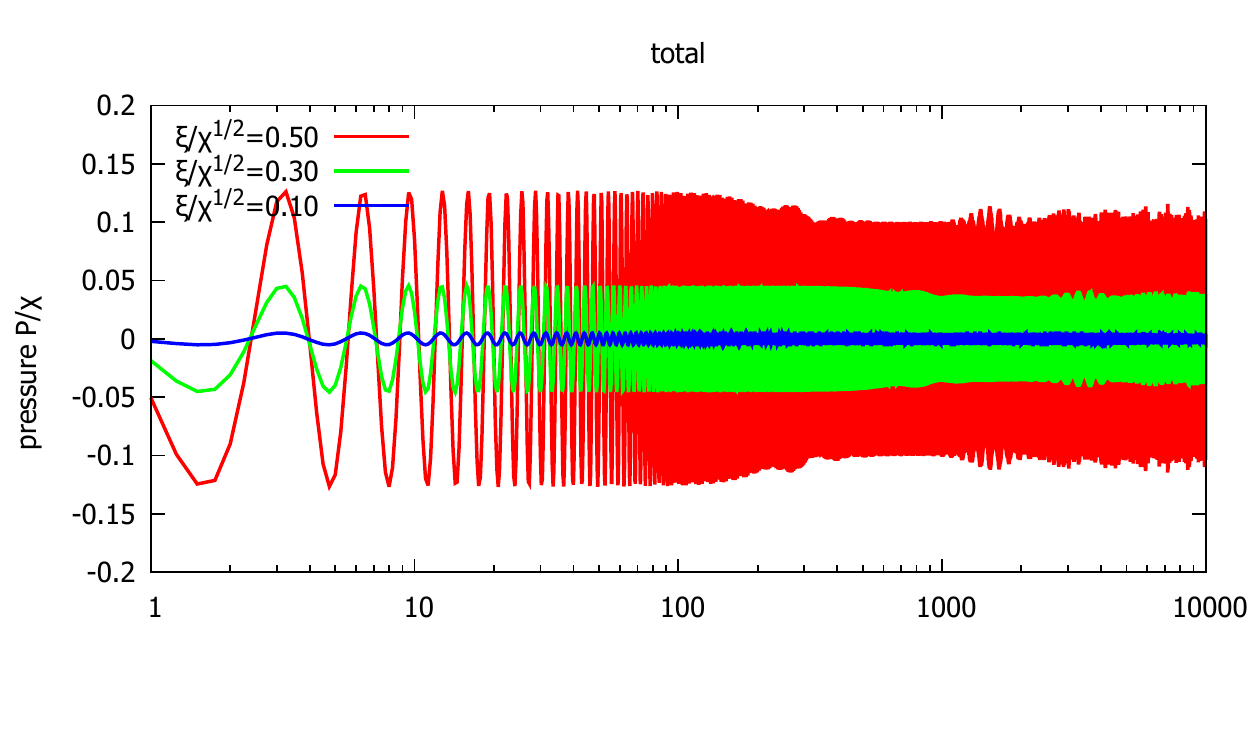} \hspace{-5.5mm}
  \includegraphics[width=0.355\textwidth]{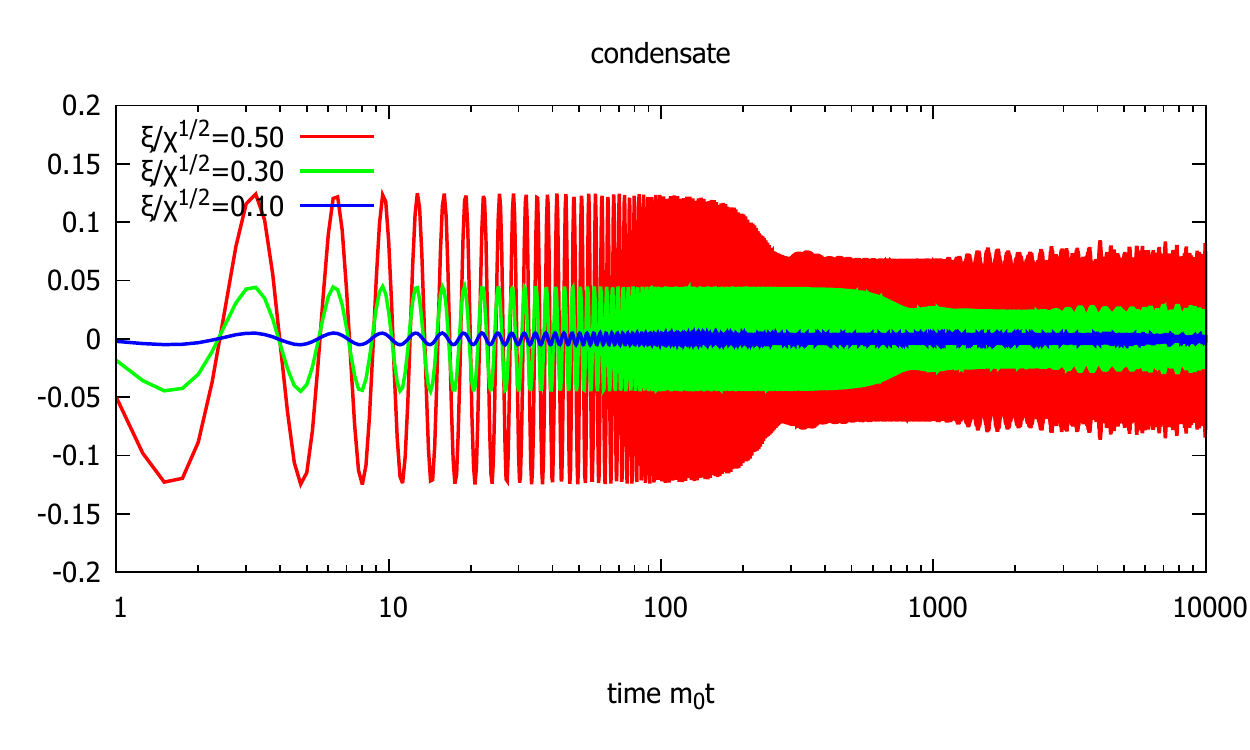} \hspace{-5.5mm}
  \includegraphics[width=0.355\textwidth]{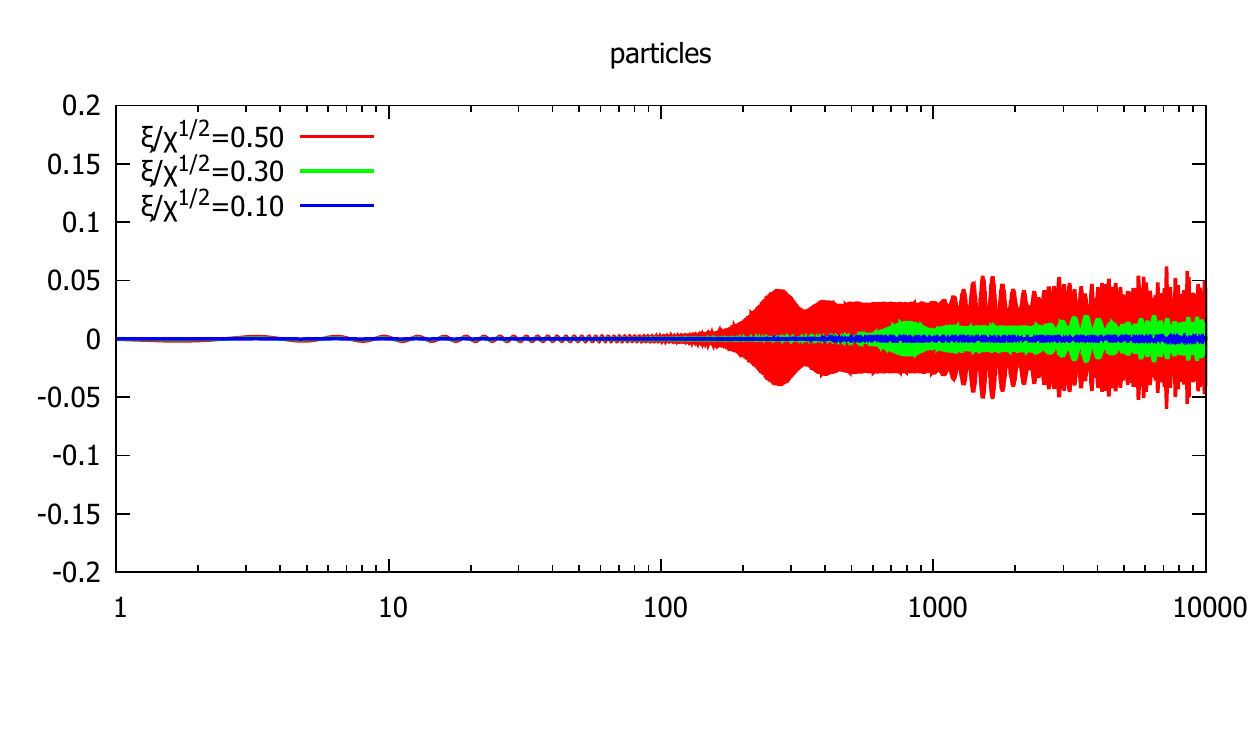} 
  \caption{Time evolution of the energy density $\braket{{\rm vac};t_{\rm in}|: \hat{\mathcal E} :|{\rm vac};t_{\rm in}}$ (top) and the pressure $\braket{{\rm vac};t_{\rm in}|: \hat{P} :|{\rm vac};t_{\rm in}}$ (bottom) for various values of initial energy $\xi$ with backreaction.  The left panel shows the total value of the system, and the center and the right panels show the contribution from the coherent axion field and the produced axion particles, respectively. }
  \label{fig4-4}
 \end{center}
\end{figure*}

Figure~\ref{fig4-4} shows the time evolution of the energy density $\braket{{\rm vac}\;t_{\rm in}|: \hat{\mathcal E} :|{\rm vac}\;t_{\rm in}}$ and the pressure $\braket{{\rm vac}\;t_{\rm in}|: \hat{P} :|{\rm vac}\;t_{\rm in}}$.  The energy density is strictly conserved because of the backreaction.  With our parameter choice $\xi/\sqrt{\chi} = {\mathcal O}(0.1)$, about one-half of the initial energy of the coherent axion field is eventually converted to axion particles.  On the other hand, the coherent field $\bar{\phi}$ dominates the total pressure.  The produced axion particles $\hat{\varphi}$ only give a relatively small contribution because axion particles are soft.

\section{Summary and Discussion} \label{sec4}

We discussed the axion particle production from coherent axion field.  We found that the coherent axion field is unstable against quantum fluctuations and it spontaneously decays by emitting axion particles when the axion potential has more than one minimum, for which quantum tunneling between different minima takes place.

In order to study the above axion particle production mechanism, we developed a formalism which describes the real-time dynamics of the axion particle production.  Namely, we derived equations of motion for coherent axion field and dynamical axion field on top of the coherent field by employing the $\hbar$-expansion and mean field approximation.  The equations are coupled with each other when axion potential is higher than quadratic, i.e., axion potential has more than one minimum.  Because of this coupling, the dynamical axion field obtains energy from the coherent field, which eventually results in the axion particle production.  We included not only $\hbar^1$-terms, but also the next leading order $\hbar^2$-term in the equations of motion in order to take into account backreaction by the axion particle production to the coherent axion field.  The backreaction is essential for the energy conservation of the system and for describing the decaying dynamics of the coherent field.

We solved the equations of motion numerically to trace the real-time dynamics of the axion particle production quantitatively.  By adopting the adiabatic particle picture and canonically quantizing the dynamical axion field at each instant of time, we evaluated the phase-space density of the produced axion particles as a function of time.  We also computed the energy density and the pressure of the system, and followed their time evolution.  As a result, we found, in particular, that (i) soft axion particles are produced from the coherent axion field, whose number grows exponentially until the production stops because of the energy conservation; (ii) the production number is much more abundant than usual perturbative axion production mechanisms, which are suppressed by the large decay constant $f_a$, and, roughly, one half of the initial energy of the coherent axion field is eventually converted into axion particles; (iii) not only axion particle production, but also an axion particle pair annihilation process occurs due to the backreaction, which leaves a characteristic striped pattern in the phase-space density; (iv) the coherent axion field loses its energy through the axion particle production, but does not completely fall into a minimum because of the pair annihilation process; and (v) inclusion of the backreaction term is important, without which the system evolves unphysically and various physical quantities including the energy density diverge.  In addition, we also compared our results with the perturbation theory.  We found that our axion particle production mechanism is genuinely nonperturbative, which cannot be described within the perturbation theory even if the initial energy of the coherent field is very small.  This is because the quantum tunneling is the physical origin of our production mechanism.

There are several interesting applications of the present study.  One example is an application to cosmology, in particular, to axion dark matter scenario.  In the standard axion dark matter scenario, the misalignment mechanism \cite{abb83, pre83, mic83} is expected to produce coherent axion field in the early Universe, whose energy density can be comparable to that of dark matter in the present Universe.  Since coherent fields have vanishing momentum, it would be a good candidate for cold dark matter.  According to our results, coherent axion field (or axion condensate) is unstable against quantum fluctuations and it decays quickly by producing ``warm'' axion particles, whose momentum is not so large $|{\bm p}| \lesssim m_0$ but nonvanishing.  In addition, the momentum distribution has a characteristic striped pattern due to the annihilation process.  These features should affect the small-scale structure of the present Universe, which might be tested by future experiments/observations.  Note that we considered a non-expanding system in this paper for simplicity.  Inclusion of expansion is important to make quantitative prediction for cosmology because the production process should be affected significantly by expansion (e.g. red-shift in momentum distribution).

Ultra-relativistic heavy-ion collisions are another example.  This is because heavy-ion collisions can produce sizable number of QCD sphalerons.  QCD sphalerons would supply energy of the order of the QCD energy scale $\Lambda_{\rm QCD}$ to coherent axion field, which in turn decays into axion particles through our production mechanism.  As an example, let us consider Pb-Pb collisions at the LHC.  In each heavy-ion collision, a quark-gluon plasma (QGP) with temperature $T \sim 200\;{\rm MeV}$, volume $V \sim 10^4\;{\rm fm}^3$, and lifetime $\tau \sim 10\;{\rm fm}$ is produced.  As the QCD sphaleron rate $\Gamma$ is given by $\Gamma \sim 30 \alpha_{\rm s}^4 T^4$ \cite{moo11}, we may estimate the number of sphalerons in a QGP created in a heavy-ion collision as $\Gamma V \tau \sim 10^4$ for $\alpha_{\rm s} \sim 0.3$.  The rate of collisions is of the order $1\;{\rm kHz}$, and thus, roughly, $10^7$ sphalerons would be produced per second in heavy-ion collisions.  On the other hand, we found that about one-half of the energy of coherent axion field is eventually converted into axion particles (see Fig.~\ref{fig4-4}).  Therefore, one QCD sphaleron may produce $0.5 \times \Lambda_{\rm QCD}/m_0 \sim 10^{+12 \sim +15}$ very soft axions.   Combining the sphaleron production number, we estimate that $N \sim 10^{+19 \sim +22}$ axion particles may be produced in a second in heavy-ion collisions.  This is a very large number compared to usual perturbative axion production processes (e.g., solar axion flux on the Earth is the order of $10^9\;{\rm cm}^2{\rm s}^{-1}$ \cite{avi17}), which is always suppressed strongly by the large decay constant $f_a$.  Therefore, heavy-ion collisions may be another good source of axion particles.  Note, however, that spatial homogeneity is assumed in the present paper.  Since a QCD sphaleron is strongly localized in space with the QCD energy scale, it is very important to consider spatial inhomogeneity to make a more realistic estimation.

It is also interesting to study the emission of gravitational waves induced by the present axion particle production mechanism.  Emission of gravitational waves by production of axion particles via parametric resonance was previously discussed in Refs.~\cite{sod18, kia18, hay19}.  We expect a similar gravitational wave emission occurs from our production mechanism as well.  If this is the case, it may leave a characteristic signature in gravitational waves due to the oscillating pattern in the momentum spectrum.  This could serve as a novel way to detect (although indirect) axion particles through observation of gravitational waves.  We leave these topics for a future work.

\section*{Acknowledgments}

H.~T. would like to thank RIKEN iTHEMS STAMP working group for useful discussions, and is supported by National Natural Science Foundation in China (NSFC) under Grant No.~11847206. X.-G.~H is supported by NSFC under Grants No.~11535012 and No.~11675041.  D.~K. is supported by the U.S. Department of Energy, Office of Nuclear Physics, under Awards No. DE-FG-88ER40388 and No. DE-AC02-98CH10886, and by the Office of Basic Energy Science under Contract No. DE-SC-0017662.

\appendix
\section{Perturbative axion production without backreaction} \label{appa}

In this appendix, we analytically discuss the axion particle production from coherent axion field within the lowest order perturbation theory by neglecting the backreaction effect.

To this end, we consider a Lagrangian for dynamical axion field $\hat{\varphi}$ given by
\begin{align}
	{\mathcal L}_{\varphi} = - \frac{1}{2} \hat{\varphi} \left[ \partial^{\mu}\partial_{\mu} + V''[\bar{\phi}]   \right] \hat{ \varphi},    \label{eqa1}
\end{align}
where ${\mathcal O}(\hbar^{3/2})$ terms are neglected (see Eq.~(\ref{eq6b})).  Here, the coherent field $\bar{\phi}$ obeys the equation of motion (\ref{eq6a}) without the backreaction term, i.e.,
\begin{align}
	0 = \partial_{\mu}\partial^{\mu} \bar{\phi} + V'[\bar{\phi}] .
\end{align}

Now, let us assume that the coherent field $\bar{\phi}$ does not deviate significantly from a minimum of the axion potential $V$, which we write $\bar{\phi}_0$.  Physically, this assumption is equivalent to assume, as we shall see explicitly later, that the initial energy of the coherent field is sufficiently smaller than the height of the potential.  We, then, perturbatively expand the Lagrangian (\ref{eqa1}) in terms of
\begin{align}
	\delta \bar{\phi} \equiv \bar{\phi} - \bar{\phi}_0
\end{align}
as
\begin{align}
	{\mathcal L}_{\varphi}
		&= - \frac{1}{2} \hat{\varphi} \left[ \partial_{\mu} \partial^{\mu} + m^2 \right] \hat{\varphi} - \frac{1}{2} \sum_{n=1}^{\infty} \frac{V^{(n+2)}}{n!} \delta \bar{\phi}^n \hat{\varphi}^2 \nonumber\\
		&\equiv {\mathcal L}_0 + \sum_{n=1}^{\infty} {\mathcal L}_n 	
		\equiv {\mathcal L}_0 + {\mathcal L}_{\rm int}.  \label{eqa5}
\end{align}
Here,
\begin{align}
	m^2 \equiv \left. \frac{\partial^2 V}{\partial \bar{\phi}^2} \right|_{\bar{\phi}=\bar{\phi}_0}, \ \
	V^{(n+2)} \equiv \left. \frac{\partial^{n+2} V}{\partial \bar{\phi}^{n+2}} \right|_{\bar{\phi}=\bar{\phi}_0}.
\end{align}

In the standard $S$-matrix formalism, the $S$-matrix is defined as a time-ordered product of the interaction Lagrangian ${\mathcal L}_{\rm int}$ as
\begin{align}
	S = {\sf T} \exp \left[  -i\int d^4x {\mathcal L}_{\rm int}  \right].
\end{align}
Then, the in- and out-state annihilation operators are related with each other by the $S$-matrix as
\begin{align}
	\hat{a}^{\rm (out)}_{\bm p}
	&= S^{\dagger} \hat{a}^{\rm (in)}_{\bm p} S \nonumber\\
	&= \hat{a}_{\bm p}^{\rm (in)} - i \int d^4 x \left[ \hat{a}_{\bm p}^{\rm (in)}, {\mathcal L}_{\rm int}  \right] + {\mathcal O}({\mathcal L}_{\rm int}^2).   \label{eqa6}
\end{align}
Therefore, in the lowest order in ${\mathcal L}_{\rm int}$, we have
\begin{align}
	\frac{d^3N^{\rm (pert)}}{d{\bm p}^3}
		&\equiv \braket{{\rm vac} | \hat{a}^{{\rm (out)}\dagger}_{\bm p}  \hat{a}^{\rm (out)}_{\bm p}   | {\rm vac}  } \nonumber\\
		&= \left| \int d^4 x \left[ \hat{a}_{\bm p}^{\rm (in)}, {\mathcal L}_{\rm int}  \right]  \right|^2.  \label{eqa7}
\end{align}

For the periodic axion potential (\ref{eq34}), the interaction Lagrangian ${\mathcal L}_{\rm int}$ starts from $n=2$, i.e.,
\begin{align}
	{\mathcal L}_{\rm int}
	&= -\frac{1}{4} \frac{m_0^2}{f_a^2} \delta\bar{\phi}^2 \hat{\varphi}^2 + {\mathcal O}(\delta\bar{\phi}^4) \nonumber\\
	&= - \frac{m_0^2}{4} \frac{\xi^2}{\chi} \sin^2(m_0 t) \hat{\varphi}^2 + {\mathcal O}\left( \frac{\xi^4}{\chi^2} \right),
\end{align}
where we used Eq.~(\ref{eq40}) to get the second line.  By plugging this expression into Eq.~(\ref{eqa7}), one obtains
\begin{align}
	\frac{1}{VT} \frac{d^3N^{\rm (pert)}}{d{\bm p}^3}
	= \frac{m_0^4}{(64\pi)^2} \left(\frac{\xi}{\sqrt{\chi} }\right)^4 \frac{\delta(|{\bm p}|)}{{\bm p}^2}.
\end{align}
The total number $N^{\rm (pert)}$ can be obtained by integrating this expression over ${\bm p}$ as
\begin{align}
	\frac{N^{\rm (pert)}}{VT} = \frac{m_0^4}{2048\pi} \left(\frac{\xi}{\sqrt{\chi} }\right)^4 ,
\end{align}
where we used $\int_0^{\infty}d|{\bm p}| \delta(|{\bm p}|)=1/2$.

\end{document}